\RequirePackage{lineno}
\documentclass[aps,prl,twocolumn,10pt,amsmath,superscriptaddress,amssymb,longbibliography]{revtex4-2}
\usepackage[normalem]{ulem}
\usepackage{epsfig,epstopdf,dcolumn,amsbsy,bm,color,CJK,amsmath,amssymb,amsfonts,appendix}
\usepackage{graphicx,algorithm,enumerate,makeidx,braket,titlesec,lineno}
\usepackage[usenames,dvipsnames]{xcolor}
\usepackage[driverfallback=dvipdfmx,colorlinks,linkcolor=blue,citecolor=blue,urlcolor=blue]{hyperref}
\usepackage{setspace}

\begin{document}

\title{Distinguishing dual lattice by strong-pulse matter-wave diffraction}

\author{Fangde Liu}
\email[These authors contributed equally to this work.]{}
\author{Wei Han}
\email[These authors contributed equally to this work.]{}
\author{Yunda Li}
\affiliation{State Key Laboratory of Quantum Optics
Technologies and Devices, \\  Institute of Opto-Electronics, \\ Collaborative Innovation Center of Extreme Optics, Shanxi
University, Taiyuan, Shanxi 030006, People's Republic of China}
\author{Feifan Zhao}
\affiliation{State Key Laboratory of Quantum Optics
Technologies and Devices, \\  Institute of Opto-Electronics, \\ Collaborative Innovation Center of Extreme Optics, Shanxi
University, Taiyuan, Shanxi 030006, People's Republic of China}
\author{Liangchao Chen}
\affiliation{State Key Laboratory of Quantum Optics
Technologies and Devices, \\  Institute of Opto-Electronics, \\ Collaborative Innovation Center of Extreme Optics, Shanxi
University, Taiyuan, Shanxi 030006, People's Republic of China}
\affiliation{Hefei National Laboratory, Hefei, Anhui 230088, People's Republic of China}
\author{Lianghui Huang}
\affiliation{State Key Laboratory of Quantum Optics
Technologies and Devices, \\  Institute of Opto-Electronics, \\ Collaborative Innovation Center of Extreme Optics, Shanxi
University, Taiyuan, Shanxi 030006, People's Republic of China}
\affiliation{Hefei National Laboratory, Hefei, Anhui 230088, People's Republic of China}
\author{Pengjun Wang}
\affiliation{State Key Laboratory of Quantum Optics
Technologies and Devices, \\  Institute of Opto-Electronics, \\ Collaborative Innovation Center of Extreme Optics, Shanxi
University, Taiyuan, Shanxi 030006, People's Republic of China}
\affiliation{Hefei National Laboratory, Hefei, Anhui 230088, People's Republic of China}
\author{Zengming Meng}
\email[Corresponding author email: ]{zmmeng01@sxu.edu.cn}
\affiliation{State Key Laboratory of Quantum Optics
Technologies and Devices, \\  Institute of Opto-Electronics, \\ Collaborative Innovation Center of Extreme Optics, Shanxi
University, Taiyuan, Shanxi 030006, People's Republic of China}
\affiliation{Hefei National Laboratory, Hefei, Anhui 230088, People's Republic of China}
\author{Jing Zhang}
\email[Corresponding author email: ]{jzhang74@sxu.edu.cn}
\affiliation{State Key Laboratory of Quantum Optics
Technologies and Devices, \\  Institute of Opto-Electronics, \\ Collaborative Innovation Center of Extreme Optics, Shanxi
University, Taiyuan, Shanxi 030006, People's Republic of China}
\affiliation{Hefei National Laboratory, Hefei, Anhui 230088, People's Republic of China}

\date{\today }

\begin{abstract}
Dual lattices such as honeycomb and hexagonal lattices typically obey Babinet's principle in optics, which states that the expected interference patterns of two complementary diffracting objects are identical and indistinguishable, except for their overall intensity. Here, we study Kapitza--Dirac diffraction of Bose--Einstein condensates in optical lattices and find that matter waves in dual lattices obey Babinet's principle only under the condition of weak-pulse Raman--Nath regimes. In contrast, the Kapitza--Dirac matter-wave diffraction in the strong-pulse Raman--Nath regime (corresponding to the phase wrapping method we developed to generate sub-wavelength phase structures in \emph{Sci. Rep.} 10, 5870 (2020)) can break Babinet's principle and clearly resolve the distinct interference patterns of the dual honeycomb and hexagonal lattices. This method offers exceptional precision in characterizing lattice configurations and advance the study of symmetry-related phenomena, overcoming the limitations of real-space imaging.
\end{abstract}

\maketitle

Optical lattices with ultracold atoms offer a well-established experimental platform for quantum simulation of complex strongly interacting many-body systems, owing to the excellent controllability of system parameters and refined measurement techniques \cite{Rev.Mod.Phys.80.885.2008, Nat.Phys.8.267.2012, RN1450}. A variety of optical lattice geometries can be realized by interfering different sets of laser beams \cite{RN1852,RN1738}, including honeycomb \cite{RN1663}, hexagonal \cite{RN1910}, checkerboard \cite{RN1992}, quasi-honeycomb \cite{RN1671}, Kagome \cite{RN1889}, Lieb \cite{RN1884} lattices, and superlattices \cite{RN1978,RN1698}, as well as more complex structures such as spin-dependent optical lattices \cite{RN273,RN525,RN1991} and twisted bilayer optical lattices \cite{RN1748}. Among various lattice geometries, honeycomb and hexagonal lattices have attracted significant attention due to their unique symmetries and the rich physics they can support, such as Dirac cones \cite{RN1693}, topological edge states \cite{RN1952}, and the quantum Hall effect \cite{RN1953}.

\begin{figure*}[tb]
\includegraphics[width=7.0in]{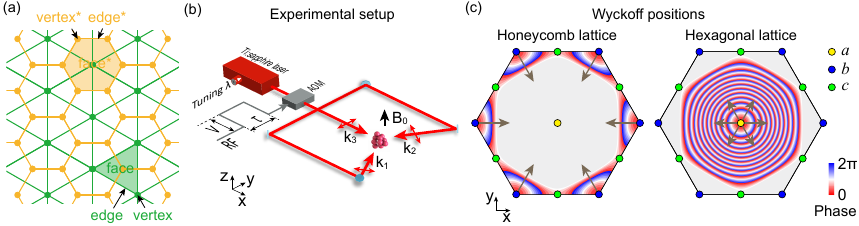}
\caption{(a) Representation of dual graph for honeycomb and hexagonal lattices. The hexagonal lattice (green) consists of vertices, edges, and faces, while its dual lattice, the honeycomb lattice (yellow), replaces each face of the hexagonal lattice with a vertex and connects vertices where adjacent faces share an edge. (b) Schematic of the experimental setup. Three coplanar lattice beams intersect at $120^{\circ}$ in the $xy$-plane with linear $p$-polarizations, while a homogeneous magnetic bias field is applied along the z-axis.
A Ti:sapphire laser is used to continuously tune the wavelength of the optical lattice. The pulse duration $\tau$ and strength $V_0$ of the lattice beams are controlled by the RF voltage $V$ applied to the acousto-optic modulator (AOM). (c) Phase wrapping of matter wave in the honeycomb and hexagonal lattices under the strong-pulse Raman--Nath regime. Atoms occupy the Wyckoff positions $b$ and $a$ in the honeycomb and hexagonal lattices, respectively. Bottom layers show the contour map of the sub-wavelength phase structures at $V_0 \tau / \hbar=12\times2\pi$. Brown arrows indicate the directions of the steepest gradient of the lattice potential.}\label{Fig1}
\end{figure*}

Optical lattices can be effectively studied using graph theory, which provides an intuitive framework for analyzing their geometric structures and revealing the underlying topological properties \cite{RN2076,RN2075}. A graph $G(v,e,f)$ is defined as a set of vertices $(v)$, edges $(e)$ and faces $(f)$, where each edge connects two vertices, and the faces represent the enclosed regions formed by these connections. The dual graph $G^*(v^*,e^*,f^*)$ of the graph $G$, formed by interchanging the roles of faces and vertices in a graph, follow the relations $v^* = f$, $e^* = e$ and $f^* = v$, where edges remain unchanged, vertices and faces are swapped between the primal and dual graphs. This duality establishes a profound connection between lattice geometries. Notably, the honeycomb lattice and hexagonal lattice exemplify this relationship  \cite{RN1997, RN1996}, as shown in Fig. \ref{Fig1}~(a). In optics, dual lattices obey Babinet’s principle, which states that the interference patterns of two complementary diffracting objects are essentially identical, except for the overall intensity of the diffracted pattern \cite{Optics, PrinciplesOfOptics}.

In this Letter, we investigate Kapitza--Dirac diffraction of matter waves in dual honeycomb and hexagonal optical lattices. We find that Babinet's principle holds in the weak-pulse Raman--Nath regime but breaks down in the strong-pulse regime, where the dual lattices exhibit markedly different diffraction patterns in momentum space. This breakdown can be attributed to the phase wrapping induced by strong pulse in the Raman--Nath regime, which encodes subwavelength real-space phase information within a unit cell into higher-order diffraction peaks in momentum space. These discrete momentum peaks self-organize into distinct diffraction patterns that reflect the different phase wrapping experienced by atoms at inequivalent Wyckoff positions in the dual lattices. Using time-of-flight imaging, we observe that the diffraction patterns of the honeycomb and hexagonal lattices in the strong-pulse Raman--Nath regime exhibit star-shaped and hexagonal structures, respectively. Furthermore, by manipulating the scalar and vector polarizabilities of the atoms, we demonstrate how to realize various lattice structures, including the dual honeycomb and hexagonal lattices, as well as more complex lattices with different symmetries, and how to distinguish them using matter-wave diffraction.

A schematic of the experimental setup is shown in Fig.~\ref{Fig1}(b). Bose--Einstein condensates (BECs) with up to $5 \times 10^5 $ Rubidium-87 ($^{87}$Rb) atoms in the spin state $\left\vert F=1,m_{F}=1\right\rangle$ of the ground-state manifold are prepared in a crossed optical dipole trap. We use three laser beams, interfered at angles of $120^\circ$ in the $xy$-plane, to create a 2D optical lattice. The polarizations of the lattice beams are linear within the plane of interference and perpendicular to a homogeneous magnetic guiding field of $B_0=2.7\,$G, which defines the quantization axis along the $z$ direction. The lattice beams are controlled by a single-pass AOM to generate optical lattice pulses. By using a Ti:sapphire laser with a continuously tunable wavelength range of $\triangle\lambda=46\,\rm{nm}$ $(774\,\rm{nm} \sim 820\,\rm{nm})$, we can flexibly manipulate the scalar and vector polarizabilities of the a.c. Stark shift, enabling the realization of tunable lattice structures ranging from honeycomb to hexagonal lattices.

The atom-laser electric dipole interaction gives rise to a lattice potential $V(\mathbf{r})$, which consists of a scalar potential $V_{\mathrm{s}}(\mathbf{r})$ and a vector potential $V_{\mathrm{v}}(\mathbf{r})$, and can be written as
\begin{eqnarray}
V\left( \mathbf{r}\right) =V_{\mathrm{s}}\left( \mathbf{r}\right)+V_{\mathrm{v}}\left( \mathbf{r}\right).
\label{Eq:Potential}
\end{eqnarray}
The scalar potential is $V_{\mathrm{s}}(\mathbf{r})=-\alpha ^{(0)}(\omega,F)|\mathbf{E}|^{2}$ and the vector potential is $V_{\mathrm{v}}( \mathbf{r})= -\alpha ^{(1)}(\omega,F) ( i\mathbf{E}^{*}\times \mathbf{E})\cdot \hat{e}_B\frac{m_F}{F}$, where $\alpha ^{\left( 0\right) }(\omega,F)$ and $\alpha ^{\left( 1\right) }(\omega,F)$ denote the scalar and vector polarizabilities, respectively~\cite{RN542}. The unit vector $\hat{e}_B$ indicates the direction of the magnetic quantization axis, $m_F$ and $F$ are the magnetic and total angular momentum quantum numbers of atoms in a hyperfine ground-state manifold. The optical electric field vector formed by the three laser beams is given by $\mathbf{E}=\sum\nolimits_{j=1}^{3}E_0\hat{\mathbf{\epsilon}}_j e^{i( \mathbf{k}_{j}\cdot \mathbf{r}-\omega t) }$ with $E_0$ denoting the field amplitude, $\hat{\mathbf{\epsilon}}_j$ the polarization vector, $\omega$ the angular frequency, and $\mathbf{k}_j$ the wave vector of the $\textit{j}$-th laser beam. For polarization perpendicular to the quantization axis $z$, the scalar potential is $V_{\mathrm{s}}(\mathbf{r})=-\alpha ^{(0)}(\omega,F)E_0^{2}\big[ 3-\sum\nolimits_{j=1}^{3}\cos( \mathbf{b}_{j}\cdot \mathbf{r})\big]$ and the vector potential is $V_{\mathrm{v}}( \mathbf{r})=-\alpha ^{(1)}(\omega,F)E_0^{2}\big[\sqrt{3}\sum\nolimits_{j=1}^{3}\sin( \mathbf{b}_{j}\cdot \mathbf{r})\big]\frac{m_F}{F}$, where $\mathbf{b}_1 = \mathbf{k}_2-\mathbf{k}_3$, $\mathbf{b}_2 =\mathbf{k}_3-\mathbf{k}_1$ and $\mathbf{b}_3 = \mathbf{k}_1-\mathbf{k}_2$ are the reciprocal lattice vectors. In the case of polarization along the quantization axis $z$, the vector potential vanishes, leaving only the scalar potential.

The lattice potential $V(\mathbf{r})$ can be effectively controlled by tuning the frequency to change the scalar and vector polarizabilities $\alpha ^{\left( 0\right) }$ and $\alpha ^{\left( 1\right) }$. A pure scalar potential $V_{\mathrm{s}}(\mathbf{r})$ with $\alpha^{(0)}>0$ (red detuning) forms a honeycomb lattice, whereas $\alpha^{(0)}<0$ (blue detuning) leads to a hexagonal lattice. A pure vector potential $V_{\mathrm{v}}(\mathbf{r})$ instead gives rise to a triangular lattice (see Fig. S1 in the Supplementary Material). When $\alpha^{(0)}=\pm \alpha^{(1)}\frac{m_F}{F}>0$, the total lattice potential $V(\mathbf{r})$ forms a hexagonal lattice. Conversely, when $\alpha^{(0)}=\pm \alpha^{(1)}\frac{m_F}{F}<0$, the lattice potential $V(\mathbf{r})$ forms a honeycomb lattice. The honeycomb and hexagonal lattice potentials, differing only by a sign, constitute a pair of dual lattices \cite{RN1997, RN1996}, as illustrated in Fig.~\ref{Fig1}(a).

\begin{figure*}[tb]
\includegraphics[width=7in]{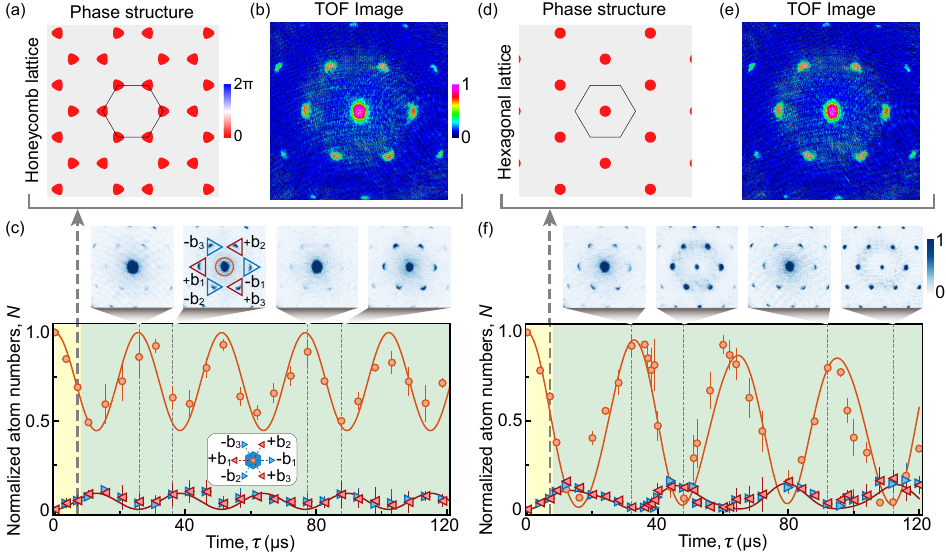}
\caption{Kapitza--Dirac diffraction of a BEC in the dual lattices with (a)-(c) the honeycomb lattice and (d)-(f) the hexagonal lattice under the weak-pulse regime, which obeys Babinet’s principle. (a) and (d) show the phase structure around the atomic Wyckoff positions for the dual lattices at $V_{0} \tau / \hbar = 0.6 \times 2 \pi$. (b) and (e) show TOF images of Kapitza--Dirac diffraction under a weak pulse of duration $\tau = 7 \, \mu s$ in the Raman--Nath regime. (c) and (f) show the time evolution of normalized atomic populations in different momentum components during Kapitza--Dirac diffraction, with the insets displaying TOF images at various pulse durations. The orange circle and line correspond the component of zero momentum, while the red (blue) triangle and line correspond to the components of the first-order diffraction momentum at $+\mathbf{b}_{1,2,3}$ ($-\mathbf{b}_{1,2,3}$). The yellow-shaded and the green-shaded regions mark the Raman--Nath and non Raman--Nath regimes, respectively. The lattice depth is fixed at $V_{0} = 23 \, E_r$.}\label{Fig2}
\end{figure*}

To characterize the lattice structure, we use Kapitza--Dirac matter-wave diffraction \cite{RN730,RN1889}. In the Raman--Nath regime, defined by a short pulse duration $\tau \ll \tau_{\text{lim}} = h/\sqrt{4V_0 E_r}$ (with $E_r$ denoting the photon recoil energy, $V_0$ the lattice depth, and $h$ Planck’s constant), the atomic kinetic energy can be neglected \cite{RN1926}. In this regime, the lattice potential $V(\mathbf{r})$ acts as a spatially dependent phase grating on the matter wave, and the evolution of the BEC can be described classically by $\Psi(\mathbf{r},\tau) = \Psi(\mathbf{r},0) e^{i V(\mathbf{r}) \tau / \hbar}$, where $\Psi(\mathbf{r},0)$ is the initial wave function. The induced phase modulation scale is directly determined by the product of the lattice depth \( V_0 \) and the pulse duration \( \tau \). While this modulation is encoded in position space, it can be revealed in momentum space through time-of-flight (TOF) absorption imaging~\cite{RN905}. Outside the Raman--Nath regime ($\tau>\tau_{\mathrm{lim}}$), the atomic kinetic energy becomes significant, rendering the phase grating approximation inadequate for accurately describing the evolution of the BEC. In this regime, the atomic kinetic and potential energies exchange periodically, leading to oscillations of atoms between low-order and high-order diffraction momentum states. Accurate determination of the diffraction dynamics requires the full quantum evolution by solving the Schr\"{o}dinger equation (see Sec.~I in the Supplementary Material).

Furthermore, the Raman--Nath regime can be divided into weak-pulse and strong-pulse regions (see Table 1 in the Supplementary Material). In the weak-pulse Raman--Nath regime, characterized by $V_0\tau_{\mathrm{lim}} / \hbar \lesssim 2\pi$, atoms predominantly populate the zeroth- and first-order diffraction momentum states. In this case, the dominant first-order diffraction peaks reflect the reciprocal lattice vectors, and thus encode the periodicity of the optical lattice. Since the dual honeycomb and hexagonal lattice potentials, differing only by a sign, share the same periodic structure, their diffraction patterns are indistinguishable in this regime. When $V_0\tau_{\mathrm{lim}} / \hbar \gg 2\pi$, the system enters the strong-pulse Raman--Nath regime. By further adjusting the parameters such that $V_0 \tau / \hbar > 2\pi$, phase wrapping emerges, leading to a subwavelength periodic modulation imprinted onto the real-space wavefunction, as shown in Fig.~\ref{Fig1}(c). This real-space modulation is encoded in the momentum distribution through the appearance of higher-order diffraction peaks. The accessible diffraction order is determined by the local gradient of the lattice potential experienced by the atoms. A steep gradient at the atomic position leads to more rapid spatial phase variation, thereby enabling diffraction into higher-order momentum states through enhanced phase wrapping. The spatial variation of the potential gradient within a primitive cell encodes the fine structure of the lattice potential into the momentum-space diffraction pattern via this phase-wrapping mechanism. For the honeycomb lattice, atoms occupy the Wyckoff position $b$, where the steepest gradients of the lattice potential predominantly align along six specific directions, as indicated by the arrows in the left panel of Fig.~\ref{Fig1}(c). Consequently, high-order momentum states are accessible exclusively along these six directions, and a characteristic star-shaped diffraction pattern is expected. In contrast, for the hexagonal lattice, atoms occupy the Wyckoff position $a$, around which the lattice potential gradient exhibits continuous rotational symmetry, as indicated by the arrows in the right panel of Fig.~\ref{Fig1}(c). This symmetry results in diffraction with uniform intensity in all directions, and a regular hexagonal diffraction pattern is expected. In optical systems, the strong-pule Raman--Nath regime remains inaccessible, thus the diffraction patterns of the dual honeycomb and hexagonal lattices exhibit no discernible differences.

In the experiment, we apply a lattice potential pulse to the BEC in the hyperfine spin state $\left\vert 1,1 \right\rangle$ for a duration $\tau$. Immediately after the pulse, the optical dipole trap is turned off, and the atoms undergo 12 ms of ballistic expansion before being detected via TOF absorption imaging. We first observe atomic diffraction from the dual lattices with a weak pulse. For pulse durations $\tau$ within the Raman--Nath regime, no phase wrapping occurs, and diffraction predominantly occurs at the first-order momentum components $\pm\textbf{b}_i$ $(i=1,2,3)$. In this regime, the dual honeycomb and hexagonal lattices are indistinguishable in their momentum diffraction patterns, as shown in Figs.~\ref{Fig2}(a)-\ref{Fig2}(b) and \ref{Fig2}(d)–\ref{Fig2}(e). For pulse durations $\tau$ outside the Raman--Nath regime, while the diffraction remains primarily at first-order momenta, the system exhibits a periodic exchange between kinetic and potential energy, leading to oscillations of atomic populations between the zeroth- and first-order diffraction momentum states, as illustrated in Figs.~\ref{Fig2}(c) and \ref{Fig2}(f). Even in this regime, the diffraction dynamics of the dual lattices remain nearly identical and indistinguishable aside from a difference in average atom number. Consequently, in the weak-pulse regime, the momentum diffraction of the dual lattices follows Babinet's principle.

\begin{figure}[tb]
\includegraphics[width=3.35in]{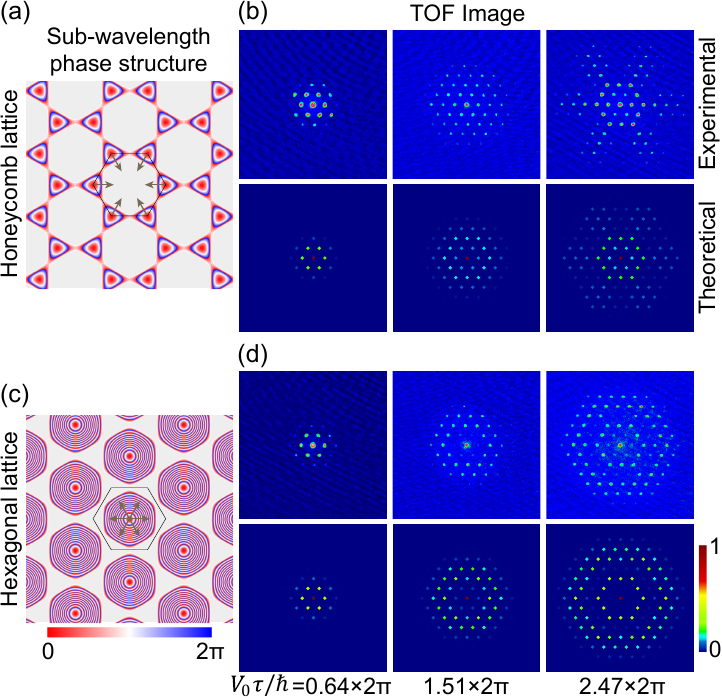}
\caption{Kapitza--Dirac diffraction of a BEC in the dual lattices with (a), (b) the honeycomb lattice and (c), (d) the hexagonal lattice under the strong-pulse Raman--Nath regime, which breaks Babinet's principle. (a), (c) Phase wrapping around the atomic Wyckoff positions of the dual lattices at $V_0 \tau / \hbar = 12 \times 2\pi$. (b), (d) Kapitza–Dirac diffraction as a function of pulse strength $V_0$ at a fixed pulse duration of $\tau = 7 \, \mu\text{s}$ in the Raman–Nath regime. The upper panels show the experimental results, while the lower panels present the theoretical predictions. From left to right, the pulse strengths are $V_0 = 25\,E_r$, $59\,E_r$, and $95\,E_r$, corresponding to phase wrapping with $V_0 \tau /\hbar =0.64\times 2\pi$, $1.51\times 2\pi$, and $2.47\times 2\pi$, respectively.}\label{Fig3}
\end{figure}

We then observe atomic diffraction from the dual lattices by fixing the pulse duration $\tau$ within the Raman--Nath regime and varying the lattice depth from weak to strong pulses. At a relatively weak lattice depth and under the condition $V_0 \tau / \hbar \ll 2\pi$, the resulting diffraction patterns are nearly identical for the honeycomb and hexagonal lattices, making them difficult to distinguish, as shown in the first column of Figs.~\ref{Fig3}(b) and \ref{Fig3}(d). As the lattice depth increases toward $V_0 \tau / \hbar \simeq 2\pi$, phase wrapping gradually emerges, the diffraction patterns formed by higher-order momentum states begin to diverge between the honeycomb and hexagonal lattices, as shown in the second column of Figs.~\ref{Fig3}(b) and \ref{Fig3}(d). In the strong-pulse regime and under the condition $V_0 \tau / \hbar \gg 2\pi$, phase wrapping becomes more pronounced, as illustrated in Figs.~\ref{Fig3}(a) and \ref{Fig3}(c). The subwavelength phase information is encoded in higher-order diffraction momentum components. Due to atoms occupying different Wyckoff positions in the dual lattices, they experience distinct local phase gradients and wrapping effects. As a result, the diffraction patterns become clearly distinguishable. This is evident in the third column of Figs.~\ref{Fig3}(b) and \ref{Fig3}(d), where the honeycomb lattice produces a star-shaped pattern due to sharp potential gradients along six discrete directions, while the hexagonal lattice yields a regular hexagonal pattern due to the continuous rotational symmetry of the potential gradients around the atomic Wyckoff position. The experimental results, obtained from TOF absorption images (first row of Figs.~\ref{Fig3}(b) and \ref{Fig3}(d)), are in excellent agreement with numerical simulations based on the full quantum evolution of the BEC (second row of Figs.~\ref{Fig3}(b) and \ref{Fig3}(d)).

\begin{figure}[tb]
\includegraphics[width=3.4in]{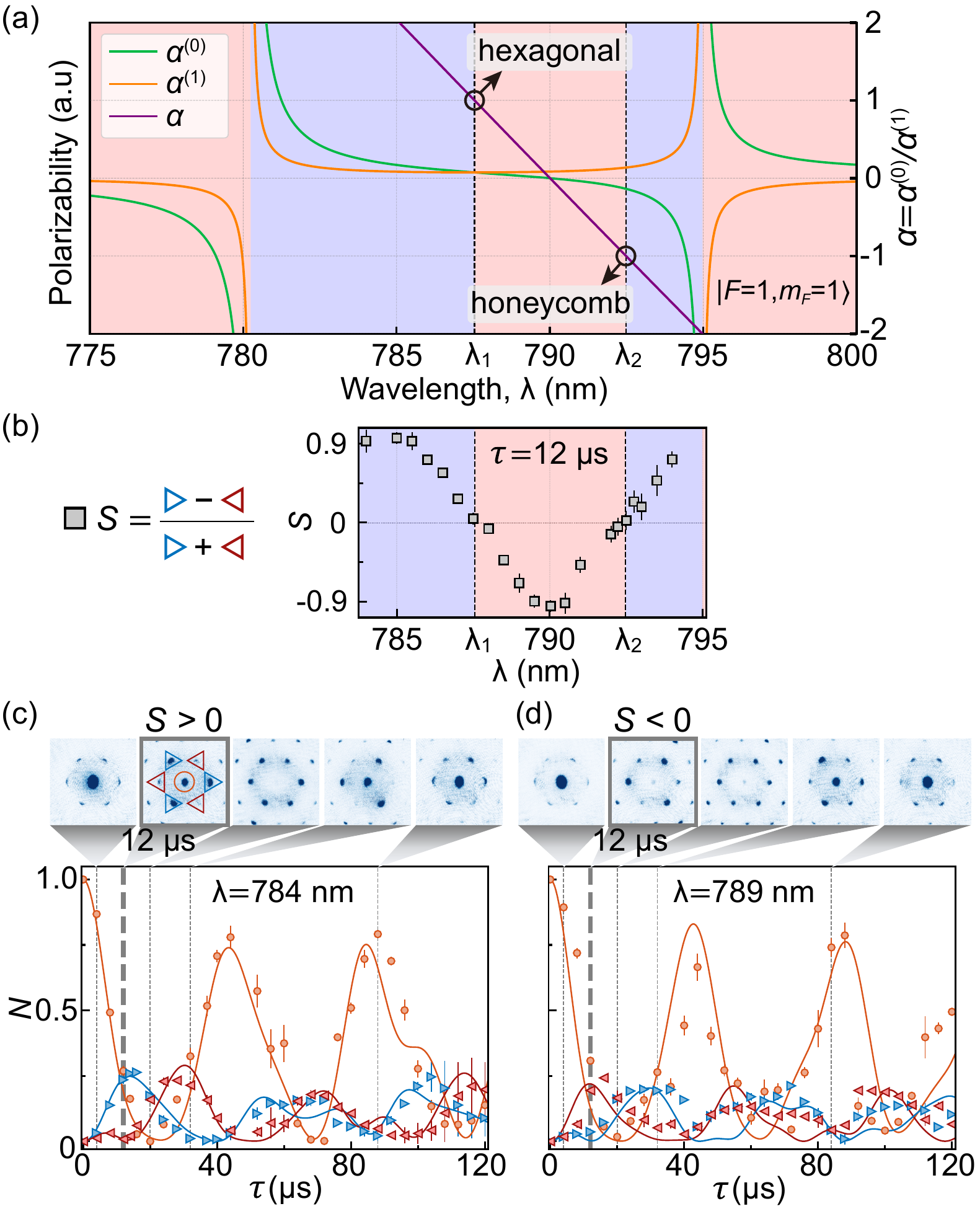}
\caption{Continuous control of the lattice structure by tuning the wavelength of the laser beams. (a) The polarizability as a function of wavelength for $^{87}$Rb atoms in the hyperfine ground state $\left\vert F = 1, m_F = 1 \right\rangle$. The ratio of scalar to vector polarizability, with $\alpha = 1$ at $\lambda_1 = 787.54\,\text{nm}$ and $\alpha = -1$ at $\lambda_2 = 792.48\,\text{nm}$, corresponds to the hexagonal and honeycomb lattices, respectively. (b) The asymmetry parameter $S$ as a function of wavelength under a weak pulse of duration $\tau = 12\,\mu s$. The region with $S > 0$ is shaded blue and that with $S < 0$ is shaded pink in (a) and (b). (c), (d) Time evolution of normalized atomic populations in different momentum components during Kapitza–Dirac diffraction for the $C_3$ rotational symmetry lattice, with $\lambda = 784$ nm in (c) and $\lambda = 789$ nm in (d). The insets show the TOF images at various pulse durations. The lattice depth is fixed at $V_{0} = 23 \, E_r$.}\label{Fig4}
\end{figure}

The emergence of phase wrapping in the strong-pulse Raman--Nath regime sheds light on the intrinsic structure and symmetry distinctions of the dual lattices. Matter-wave diffraction in the strong-pulse Raman--Nath regime breaks Babinet's principle, enabling clear resolution of the dual lattice structures in momentum space. Such a distinction is challenging to achieve in optical systems, where the strong-pulse Raman--Nath regime remains inaccessible. It is worth noting that phase wrapping has also proven highly effective in revealing underlying structural and dynamical properties, and is widely applied in fields such as structural reconstruction \cite{RN2080, RN1972, RN1973, RN1974}.

By tuning the wavelength of the lattice beams to adjust the polarizability ratio $\alpha = \alpha^{(0)}/\alpha^{(1)}$ of the atoms in the $\left\vert 1, 1\right\rangle$ state, we can achieve continuous control over the lattice structure, as shown in Fig.~\ref{Fig4}(a). When $\alpha = 1$ and $\alpha = -1$, the system realizes a hexagonal lattice and a honeycomb lattice, respectively, both exhibiting $C_6$ rotational symmetry. When $\alpha\neq\pm 1$, the lattice retains only $C_3$ rotational symmetry. The reduction of the lattice rotational symmetry from $C_6$ to $C_3$ directly leads to the opening of an energy gap at the Dirac point, enabling topological control over the energy band structure (see Fig. S2 in the Supplemental Material). To identify the resulting lattice geometry, we introduce the asymmetry parameter $S$ \cite{RN1907}, defined as $S = \frac{\sum_i( P_{\textbf{b}_{i}}-P_{-\textbf{b}_{i}}) }{\sum_i ( P_{\textbf{b}_{i}}+P_{-\textbf{b}_{i}})}$, where $P_{\textbf{b}_{i}}$ and $P_{-\textbf{b}_{i}}$ are the normalized atomic populations in the first-order diffraction peaks at $\textbf{b}_{i}$ and $-\textbf{b}_{i}$, respectively. For a weak pulse with a duration of $\tau = 12,\mu\mathrm{s}$, the asymmetry parameter $S$ as a function of the wavelength is shown in Fig.~\ref{Fig4}(b), where $S = 0$ precisely identifies the positions of the hexagonal and honeycomb lattices at $\lambda_1 = 787.54$ nm and $\lambda_2 = 792.48$ nm, respectively. It is also found that while the time evolution of atomic populations at the first-order diffraction momenta $\pm \mathbf{b}_i$ is synchronized (with $S=0$) in the hexagonal and honeycomb lattices with $C_6$ rotational symmetry, the populations at $+ \mathbf{b}_i$ and $-\mathbf{b}_i$ exhibit unsynchronized oscillations in lattices with $C_3$ rotational symmetry, as shown in Figs.~\ref{Fig4}(c) and \ref{Fig4}(d).

In conclusion, we demonstrate that Kapitza--Dirac matter-wave diffraction in the strong-pulse Raman--Nath regime breaks Babinet's principle and enables the clear resolution of dual honeycomb and hexagonal lattices in momentum space. By contrast, in the weak-pulse Raman--Nath regime, the diffraction patterns of the dual lattices remain indistinguishable, consistent with Babinet's principle. The strong-pulse Raman--Nath regime induces phase wrapping of the matter wave, which encodes real-space subwavelength phase information within a unit cell into higher-order diffraction peaks in momentum space. Atoms occupying distinct Wyckoff positions in the dual lattices experience fundamentally different phase wrapping, resulting in clearly distinguishable diffraction patterns. While the honeycomb lattice yields a star-shaped diffraction pattern, the hexagonal lattice displays a regular hexagonal pattern. By precisely adjusting the scalar and vector polarizabilities of the atoms, we achieve controllable transitions among honeycomb, hexagonal, and $C_3$-rotationally symmetric lattice geometries. This work establishes strong-pulse Kapitza--Dirac diffraction as a sensitive probe that reveals unique momentum-space fingerprints of the underlying lattice fine structure, offering new opportunities for precision lattice characterization and engineering in ultracold atomic systems.

\begin{acknowledgments}
This research is supported by the National Key Research and Development Program of China (Grant No. 2022YFA1404101), the Innovation Program for Quantum Science and Technology (Grant No. 2021ZD0302003), the National Natural Science Foundation of China (Grant Nos. 12488301, 12034011, U23A6004, 12322409, 12474252, 12474266, 12374245), and Tencent (Xplorer Prize).
\end{acknowledgments}

\bibliography{references}

\begin{thebibliography}{36}%
\makeatletter
\providecommand \@ifxundefined [1]{%
 \@ifx{#1\undefined}
}%
\providecommand \@ifnum [1]{%
 \ifnum #1\expandafter \@firstoftwo
 \else \expandafter \@secondoftwo
 \fi
}%
\providecommand \@ifx [1]{%
 \ifx #1\expandafter \@firstoftwo
 \else \expandafter \@secondoftwo
 \fi
}%
\providecommand \natexlab [1]{#1}%
\providecommand \enquote  [1]{``#1''}%
\providecommand \bibnamefont  [1]{#1}%
\providecommand \bibfnamefont [1]{#1}%
\providecommand \citenamefont [1]{#1}%
\providecommand \href@noop [0]{\@secondoftwo}%
\providecommand \href [0]{\begingroup \@sanitize@url \@href}%
\providecommand \@href[1]{\@@startlink{#1}\@@href}%
\providecommand \@@href[1]{\endgroup#1\@@endlink}%
\providecommand \@sanitize@url [0]{\catcode `\\12\catcode `\$12\catcode
  `\&12\catcode `\#12\catcode `\^12\catcode `\_12\catcode `\%12\relax}%
\providecommand \@@startlink[1]{}%
\providecommand \@@endlink[0]{}%
\providecommand \url  [0]{\begingroup\@sanitize@url \@url }%
\providecommand \@url [1]{\endgroup\@href {#1}{\urlprefix }}%
\providecommand \urlprefix  [0]{URL }%
\providecommand \Eprint [0]{\href }%
\providecommand \doibase [0]{https://doi.org/}%
\providecommand \selectlanguage [0]{\@gobble}%
\providecommand \bibinfo  [0]{\@secondoftwo}%
\providecommand \bibfield  [0]{\@secondoftwo}%
\providecommand \translation [1]{[#1]}%
\providecommand \BibitemOpen [0]{}%
\providecommand \bibitemStop [0]{}%
\providecommand \bibitemNoStop [0]{.\EOS\space}%
\providecommand \EOS [0]{\spacefactor3000\relax}%
\providecommand \BibitemShut  [1]{\csname bibitem#1\endcsname}%
\let\auto@bib@innerbib\@empty
\bibitem [{\citenamefont {Bloch}\ \emph {et~al.}(2008)\citenamefont {Bloch},
  \citenamefont {Dalibard},\ and\ \citenamefont
  {Zwerger}}]{Rev.Mod.Phys.80.885.2008}%
  \BibitemOpen
  \bibfield  {author} {\bibinfo {author} {\bibfnamefont {I.}~\bibnamefont
  {Bloch}}, \bibinfo {author} {\bibfnamefont {J.}~\bibnamefont {Dalibard}},\
  and\ \bibinfo {author} {\bibfnamefont {W.}~\bibnamefont {Zwerger}},\
  }\bibfield  {title} {\bibinfo {title} {Many-body physics with ultracold
  gases},\ }\href {https://doi.org/10.1103/RevModPhys.80.885} {\bibfield
  {journal} {\bibinfo  {journal} {Rev. Mod. Phys.}\ }\textbf {\bibinfo {volume}
  {80}},\ \bibinfo {pages} {885} (\bibinfo {year} {2008})}\BibitemShut
  {NoStop}%
\bibitem [{\citenamefont {Bloch}\ \emph {et~al.}(2012)\citenamefont {Bloch},
  \citenamefont {Dalibard},\ and\ \citenamefont
  {Nascimbène}}]{Nat.Phys.8.267.2012}%
  \BibitemOpen
  \bibfield  {author} {\bibinfo {author} {\bibfnamefont {I.}~\bibnamefont
  {Bloch}}, \bibinfo {author} {\bibfnamefont {J.}~\bibnamefont {Dalibard}},\
  and\ \bibinfo {author} {\bibfnamefont {S.}~\bibnamefont {Nascimbène}},\
  }\bibfield  {title} {\bibinfo {title} {Quantum simulations with ultracold
  quantum gases},\ }\href {https://doi.org/10.1038/nphys2259} {\bibfield
  {journal} {\bibinfo  {journal} {Nat. Phys.}\ }\textbf {\bibinfo {volume}
  {8}},\ \bibinfo {pages} {267} (\bibinfo {year} {2012})}\BibitemShut {NoStop}%
\bibitem [{\citenamefont {Lewenstein}\ \emph {et~al.}(2012)\citenamefont
  {Lewenstein}, \citenamefont {Sanpera},\ and\ \citenamefont
  {Ahufinger}}]{RN1450}%
  \BibitemOpen
  \bibfield  {author} {\bibinfo {author} {\bibfnamefont {M.}~\bibnamefont
  {Lewenstein}}, \bibinfo {author} {\bibfnamefont {A.}~\bibnamefont
  {Sanpera}},\ and\ \bibinfo {author} {\bibfnamefont {V.}~\bibnamefont
  {Ahufinger}},\ }\href@noop {} {\emph {\bibinfo {title} {Ultracold Atoms in
  Optical Lattices: Simulating quantum many-body systems}}}\ (\bibinfo
  {publisher} {Oxford University Press},\ \bibinfo {year} {2012})\BibitemShut
  {NoStop}%
\bibitem [{\citenamefont {Windpassinger}\ and\ \citenamefont
  {Sengstock}(2013)}]{RN1852}%
  \BibitemOpen
  \bibfield  {author} {\bibinfo {author} {\bibfnamefont {P.}~\bibnamefont
  {Windpassinger}}\ and\ \bibinfo {author} {\bibfnamefont {K.}~\bibnamefont
  {Sengstock}},\ }\bibfield  {title} {\bibinfo {title} {Engineering novel
  optical lattices},\ }\href {https://doi.org/10.1088/0034-4885/76/8/086401}
  {\bibfield  {journal} {\bibinfo  {journal} {Rep. Prog. Phys.}\ }\textbf
  {\bibinfo {volume} {76}},\ \bibinfo {pages} {086401} (\bibinfo {year}
  {2013})}\BibitemShut {NoStop}%
\bibitem [{\citenamefont {Schäfer}\ \emph {et~al.}(2020)\citenamefont
  {Schäfer}, \citenamefont {Fukuhara}, \citenamefont {Sugawa}, \citenamefont
  {Takasu},\ and\ \citenamefont {Takahashi}}]{RN1738}%
  \BibitemOpen
  \bibfield  {author} {\bibinfo {author} {\bibfnamefont {F.}~\bibnamefont
  {Schäfer}}, \bibinfo {author} {\bibfnamefont {T.}~\bibnamefont {Fukuhara}},
  \bibinfo {author} {\bibfnamefont {S.}~\bibnamefont {Sugawa}}, \bibinfo
  {author} {\bibfnamefont {Y.}~\bibnamefont {Takasu}},\ and\ \bibinfo {author}
  {\bibfnamefont {Y.}~\bibnamefont {Takahashi}},\ }\bibfield  {title} {\bibinfo
  {title} {Tools for quantum simulation with ultracold atoms in optical
  lattices},\ }\href {https://doi.org/10.1038/s42254-020-0195-3} {\bibfield
  {journal} {\bibinfo  {journal} {Nat. Rev. Phys.}\ }\textbf {\bibinfo {volume}
  {2}},\ \bibinfo {pages} {411} (\bibinfo {year} {2020})}\BibitemShut {NoStop}%
\bibitem [{\citenamefont {Brown}\ \emph {et~al.}(2022)\citenamefont {Brown},
  \citenamefont {Chang}, \citenamefont {Schwarz}, \citenamefont {Leung},
  \citenamefont {Kozii}, \citenamefont {Avdoshkin}, \citenamefont {Moore},\
  and\ \citenamefont {Stamper-Kurn}}]{RN1663}%
  \BibitemOpen
  \bibfield  {author} {\bibinfo {author} {\bibfnamefont {C.~D.}\ \bibnamefont
  {Brown}}, \bibinfo {author} {\bibfnamefont {S.}~\bibnamefont {Chang}},
  \bibinfo {author} {\bibfnamefont {M.~N.}\ \bibnamefont {Schwarz}}, \bibinfo
  {author} {\bibfnamefont {T.~H.}\ \bibnamefont {Leung}}, \bibinfo {author}
  {\bibfnamefont {V.}~\bibnamefont {Kozii}}, \bibinfo {author} {\bibfnamefont
  {A.}~\bibnamefont {Avdoshkin}}, \bibinfo {author} {\bibfnamefont {J.~E.}\
  \bibnamefont {Moore}},\ and\ \bibinfo {author} {\bibfnamefont
  {D.}~\bibnamefont {Stamper-Kurn}},\ }\bibfield  {title} {\bibinfo {title}
  {Direct geometric probe of singularities in band structure},\ }\href
  {https://doi.org/doi:10.1126/science.abm6442} {\bibfield  {journal} {\bibinfo
   {journal} {Science}\ }\textbf {\bibinfo {volume} {377}},\ \bibinfo {pages}
  {1319} (\bibinfo {year} {2022})}\BibitemShut {NoStop}%
\bibitem [{\citenamefont {Struck}\ \emph {et~al.}(2011)\citenamefont {Struck},
  \citenamefont {Ölschläger}, \citenamefont {Le~Targat}, \citenamefont
  {Soltan-Panahi}, \citenamefont {Eckardt}, \citenamefont {Lewenstein},
  \citenamefont {Windpassinger},\ and\ \citenamefont {Sengstock}}]{RN1910}%
  \BibitemOpen
  \bibfield  {author} {\bibinfo {author} {\bibfnamefont {J.}~\bibnamefont
  {Struck}}, \bibinfo {author} {\bibfnamefont {C.}~\bibnamefont
  {Ölschläger}}, \bibinfo {author} {\bibfnamefont {R.}~\bibnamefont
  {Le~Targat}}, \bibinfo {author} {\bibfnamefont {P.}~\bibnamefont
  {Soltan-Panahi}}, \bibinfo {author} {\bibfnamefont {A.}~\bibnamefont
  {Eckardt}}, \bibinfo {author} {\bibfnamefont {M.}~\bibnamefont {Lewenstein}},
  \bibinfo {author} {\bibfnamefont {P.}~\bibnamefont {Windpassinger}},\ and\
  \bibinfo {author} {\bibfnamefont {K.}~\bibnamefont {Sengstock}},\ }\bibfield
  {title} {\bibinfo {title} {Quantum simulation of frustrated classical
  magnetism in triangular optical lattices},\ }\href
  {https://doi.org/doi:10.1126/science.1207239} {\bibfield  {journal} {\bibinfo
   {journal} {Science}\ }\textbf {\bibinfo {volume} {333}},\ \bibinfo {pages}
  {996} (\bibinfo {year} {2011})}\BibitemShut {NoStop}%
\bibitem [{\citenamefont {Sebby-Strabley}\ \emph {et~al.}(2006)\citenamefont
  {Sebby-Strabley}, \citenamefont {Anderlini}, \citenamefont {Jessen},\ and\
  \citenamefont {Porto}}]{RN1992}%
  \BibitemOpen
  \bibfield  {author} {\bibinfo {author} {\bibfnamefont {J.}~\bibnamefont
  {Sebby-Strabley}}, \bibinfo {author} {\bibfnamefont {M.}~\bibnamefont
  {Anderlini}}, \bibinfo {author} {\bibfnamefont {P.~S.}\ \bibnamefont
  {Jessen}},\ and\ \bibinfo {author} {\bibfnamefont {J.~V.}\ \bibnamefont
  {Porto}},\ }\bibfield  {title} {\bibinfo {title} {Lattice of double wells for
  manipulating pairs of cold atoms},\ }\href
  {https://doi.org/10.1103/PhysRevA.73.033605} {\bibfield  {journal} {\bibinfo
  {journal} {Phys. Rev. A}\ }\textbf {\bibinfo {volume} {73}},\ \bibinfo
  {pages} {033605} (\bibinfo {year} {2006})}\BibitemShut {NoStop}%
\bibitem [{\citenamefont {Tarruell}\ \emph {et~al.}(2012)\citenamefont
  {Tarruell}, \citenamefont {Greif}, \citenamefont {Uehlinger}, \citenamefont
  {Jotzu},\ and\ \citenamefont {Esslinger}}]{RN1671}%
  \BibitemOpen
  \bibfield  {author} {\bibinfo {author} {\bibfnamefont {L.}~\bibnamefont
  {Tarruell}}, \bibinfo {author} {\bibfnamefont {D.}~\bibnamefont {Greif}},
  \bibinfo {author} {\bibfnamefont {T.}~\bibnamefont {Uehlinger}}, \bibinfo
  {author} {\bibfnamefont {G.}~\bibnamefont {Jotzu}},\ and\ \bibinfo {author}
  {\bibfnamefont {T.}~\bibnamefont {Esslinger}},\ }\bibfield  {title} {\bibinfo
  {title} {Creating, moving and merging {Dirac points with a Fermi} gas in a
  tunable honeycomb lattice},\ }\href {https://doi.org/10.1038/nature10871}
  {\bibfield  {journal} {\bibinfo  {journal} {Nature}\ }\textbf {\bibinfo
  {volume} {483}},\ \bibinfo {pages} {302} (\bibinfo {year}
  {2012})}\BibitemShut {NoStop}%
\bibitem [{\citenamefont {Jo}\ \emph {et~al.}(2012)\citenamefont {Jo},
  \citenamefont {Guzman}, \citenamefont {Thomas}, \citenamefont {Hosur},
  \citenamefont {Vishwanath},\ and\ \citenamefont {Stamper-Kurn}}]{RN1889}%
  \BibitemOpen
  \bibfield  {author} {\bibinfo {author} {\bibfnamefont {G.~B.}\ \bibnamefont
  {Jo}}, \bibinfo {author} {\bibfnamefont {J.}~\bibnamefont {Guzman}}, \bibinfo
  {author} {\bibfnamefont {C.~K.}\ \bibnamefont {Thomas}}, \bibinfo {author}
  {\bibfnamefont {P.}~\bibnamefont {Hosur}}, \bibinfo {author} {\bibfnamefont
  {A.}~\bibnamefont {Vishwanath}},\ and\ \bibinfo {author} {\bibfnamefont
  {D.~M.}\ \bibnamefont {Stamper-Kurn}},\ }\bibfield  {title} {\bibinfo {title}
  {Ultracold atoms in a tunable optical kagome lattice},\ }\href
  {https://doi.org/10.1103/PhysRevLett.108.045305} {\bibfield  {journal}
  {\bibinfo  {journal} {Phys. Rev. Lett.}\ }\textbf {\bibinfo {volume} {108}},\
  \bibinfo {pages} {045305} (\bibinfo {year} {2012})}\BibitemShut {NoStop}%
\bibitem [{\citenamefont {Taie}\ \emph {et~al.}(2015)\citenamefont {Taie},
  \citenamefont {Ozawa}, \citenamefont {Ichinose}, \citenamefont {Nishio},
  \citenamefont {Nakajima},\ and\ \citenamefont {Takahashi}}]{RN1884}%
  \BibitemOpen
  \bibfield  {author} {\bibinfo {author} {\bibfnamefont {S.}~\bibnamefont
  {Taie}}, \bibinfo {author} {\bibfnamefont {H.}~\bibnamefont {Ozawa}},
  \bibinfo {author} {\bibfnamefont {T.}~\bibnamefont {Ichinose}}, \bibinfo
  {author} {\bibfnamefont {T.}~\bibnamefont {Nishio}}, \bibinfo {author}
  {\bibfnamefont {S.}~\bibnamefont {Nakajima}},\ and\ \bibinfo {author}
  {\bibfnamefont {Y.}~\bibnamefont {Takahashi}},\ }\bibfield  {title} {\bibinfo
  {title} {Coherent driving and freezing of bosonic matter wave in an optical
  {Lieb} lattice},\ }\href {https://doi.org/doi:10.1126/sciadv.1500854}
  {\bibfield  {journal} {\bibinfo  {journal} {Sci. Adv.}\ }\textbf {\bibinfo
  {volume} {1}},\ \bibinfo {pages} {e1500854} (\bibinfo {year}
  {2015})}\BibitemShut {NoStop}%
\bibitem [{\citenamefont {Yang}\ \emph {et~al.}(2020)\citenamefont {Yang},
  \citenamefont {Sun}, \citenamefont {Huang}, \citenamefont {Wang},
  \citenamefont {Deng}, \citenamefont {Dai}, \citenamefont {Yuan},\ and\
  \citenamefont {Pan}}]{RN1978}%
  \BibitemOpen
  \bibfield  {author} {\bibinfo {author} {\bibfnamefont {B.}~\bibnamefont
  {Yang}}, \bibinfo {author} {\bibfnamefont {H.}~\bibnamefont {Sun}}, \bibinfo
  {author} {\bibfnamefont {C.}~\bibnamefont {Huang}}, \bibinfo {author}
  {\bibfnamefont {H.}~\bibnamefont {Wang}}, \bibinfo {author} {\bibfnamefont
  {Y.}~\bibnamefont {Deng}}, \bibinfo {author} {\bibfnamefont {H.}~\bibnamefont
  {Dai}}, \bibinfo {author} {\bibfnamefont {Z.}~\bibnamefont {Yuan}},\ and\
  \bibinfo {author} {\bibfnamefont {J.}~\bibnamefont {Pan}},\ }\bibfield
  {title} {\bibinfo {title} {Cooling and entangling ultracold atoms in optical
  lattices},\ }\href {https://doi.org/doi:10.1126/science.aaz6801} {\bibfield
  {journal} {\bibinfo  {journal} {Science}\ }\textbf {\bibinfo {volume}
  {369}},\ \bibinfo {pages} {550} (\bibinfo {year} {2020})}\BibitemShut
  {NoStop}%
\bibitem [{\citenamefont {Kosch}\ \emph {et~al.}(2022)\citenamefont {Kosch},
  \citenamefont {Asteria}, \citenamefont {Zahn}, \citenamefont {Sengstock},\
  and\ \citenamefont {Weitenberg}}]{RN1698}%
  \BibitemOpen
  \bibfield  {author} {\bibinfo {author} {\bibfnamefont {M.~N.}\ \bibnamefont
  {Kosch}}, \bibinfo {author} {\bibfnamefont {L.}~\bibnamefont {Asteria}},
  \bibinfo {author} {\bibfnamefont {H.~P.}\ \bibnamefont {Zahn}}, \bibinfo
  {author} {\bibfnamefont {K.}~\bibnamefont {Sengstock}},\ and\ \bibinfo
  {author} {\bibfnamefont {C.}~\bibnamefont {Weitenberg}},\ }\bibfield  {title}
  {\bibinfo {title} {Multifrequency optical lattice for dynamic
  lattice-geometry control},\ }\href
  {https://doi.org/10.1103/PhysRevResearch.4.043083} {\bibfield  {journal}
  {\bibinfo  {journal} {Phys. Rev. Res.}\ }\textbf {\bibinfo {volume} {4}},\
  \bibinfo {pages} {043083} (\bibinfo {year} {2022})}\BibitemShut {NoStop}%
\bibitem [{\citenamefont {Mandel}\ \emph {et~al.}(2003)\citenamefont {Mandel},
  \citenamefont {Greiner}, \citenamefont {Widera}, \citenamefont {Rom},
  \citenamefont {Hänsch},\ and\ \citenamefont {Bloch}}]{RN273}%
  \BibitemOpen
  \bibfield  {author} {\bibinfo {author} {\bibfnamefont {O.}~\bibnamefont
  {Mandel}}, \bibinfo {author} {\bibfnamefont {M.}~\bibnamefont {Greiner}},
  \bibinfo {author} {\bibfnamefont {A.}~\bibnamefont {Widera}}, \bibinfo
  {author} {\bibfnamefont {T.}~\bibnamefont {Rom}}, \bibinfo {author}
  {\bibfnamefont {T.~W.}\ \bibnamefont {Hänsch}},\ and\ \bibinfo {author}
  {\bibfnamefont {I.}~\bibnamefont {Bloch}},\ }\bibfield  {title} {\bibinfo
  {title} {Coherent transport of neutral atoms in spin-dependent optical
  lattice potentials},\ }\href {https://doi.org/10.1103/PhysRevLett.91.010407}
  {\bibfield  {journal} {\bibinfo  {journal} {Phys. Rev. Lett.}\ }\textbf
  {\bibinfo {volume} {91}},\ \bibinfo {pages} {010407} (\bibinfo {year}
  {2003})}\BibitemShut {NoStop}%
\bibitem [{\citenamefont {Soltan-Panahi}\ \emph {et~al.}(2011)\citenamefont
  {Soltan-Panahi}, \citenamefont {Struck}, \citenamefont {Hauke}, \citenamefont
  {Bick}, \citenamefont {Plenkers}, \citenamefont {Meineke}, \citenamefont
  {Becker}, \citenamefont {Windpassinger}, \citenamefont {Lewenstein},\ and\
  \citenamefont {Sengstock}}]{RN525}%
  \BibitemOpen
  \bibfield  {author} {\bibinfo {author} {\bibfnamefont {P.}~\bibnamefont
  {Soltan-Panahi}}, \bibinfo {author} {\bibfnamefont {J.}~\bibnamefont
  {Struck}}, \bibinfo {author} {\bibfnamefont {P.}~\bibnamefont {Hauke}},
  \bibinfo {author} {\bibfnamefont {A.}~\bibnamefont {Bick}}, \bibinfo {author}
  {\bibfnamefont {W.}~\bibnamefont {Plenkers}}, \bibinfo {author}
  {\bibfnamefont {G.}~\bibnamefont {Meineke}}, \bibinfo {author} {\bibfnamefont
  {C.}~\bibnamefont {Becker}}, \bibinfo {author} {\bibfnamefont
  {P.}~\bibnamefont {Windpassinger}}, \bibinfo {author} {\bibfnamefont
  {M.}~\bibnamefont {Lewenstein}},\ and\ \bibinfo {author} {\bibfnamefont
  {K.}~\bibnamefont {Sengstock}},\ }\bibfield  {title} {\bibinfo {title}
  {Multi-component quantum gases in spin-dependent hexagonal lattices},\ }\href
  {https://doi.org/10.1038/nphys1916} {\bibfield  {journal} {\bibinfo
  {journal} {Nat. Phys.}\ }\textbf {\bibinfo {volume} {7}},\ \bibinfo {pages}
  {434} (\bibinfo {year} {2011})}\BibitemShut {NoStop}%
\bibitem [{\citenamefont {Soltan-Panahi}\ \emph {et~al.}(2012)\citenamefont
  {Soltan-Panahi}, \citenamefont {Lühmann}, \citenamefont {Struck},
  \citenamefont {Windpassinger},\ and\ \citenamefont {Sengstock}}]{RN1991}%
  \BibitemOpen
  \bibfield  {author} {\bibinfo {author} {\bibfnamefont {P.}~\bibnamefont
  {Soltan-Panahi}}, \bibinfo {author} {\bibfnamefont {D.~S.}\ \bibnamefont
  {Lühmann}}, \bibinfo {author} {\bibfnamefont {J.}~\bibnamefont {Struck}},
  \bibinfo {author} {\bibfnamefont {P.}~\bibnamefont {Windpassinger}},\ and\
  \bibinfo {author} {\bibfnamefont {K.}~\bibnamefont {Sengstock}},\ }\bibfield
  {title} {\bibinfo {title} {Quantum phase transition to unconventional
  multi-orbital superfluidity in optical lattices},\ }\href
  {https://doi.org/10.1038/nphys2128} {\bibfield  {journal} {\bibinfo
  {journal} {Nat. Phys.}\ }\textbf {\bibinfo {volume} {8}},\ \bibinfo {pages}
  {71} (\bibinfo {year} {2012})}\BibitemShut {NoStop}%
\bibitem [{\citenamefont {Meng}\ \emph {et~al.}(2023)\citenamefont {Meng},
  \citenamefont {Wang}, \citenamefont {Han}, \citenamefont {Liu}, \citenamefont
  {Wen}, \citenamefont {Gao}, \citenamefont {Wang}, \citenamefont {Chin},\ and\
  \citenamefont {Zhang}}]{RN1748}%
  \BibitemOpen
  \bibfield  {author} {\bibinfo {author} {\bibfnamefont {Z.}~\bibnamefont
  {Meng}}, \bibinfo {author} {\bibfnamefont {L.}~\bibnamefont {Wang}}, \bibinfo
  {author} {\bibfnamefont {W.}~\bibnamefont {Han}}, \bibinfo {author}
  {\bibfnamefont {F.}~\bibnamefont {Liu}}, \bibinfo {author} {\bibfnamefont
  {K.}~\bibnamefont {Wen}}, \bibinfo {author} {\bibfnamefont {C.}~\bibnamefont
  {Gao}}, \bibinfo {author} {\bibfnamefont {P.}~\bibnamefont {Wang}}, \bibinfo
  {author} {\bibfnamefont {C.}~\bibnamefont {Chin}},\ and\ \bibinfo {author}
  {\bibfnamefont {J.}~\bibnamefont {Zhang}},\ }\bibfield  {title} {\bibinfo
  {title} {Atomic {Bose-Einstein} condensate in twisted-bilayer optical
  lattices},\ }\href {https://doi.org/10.1038/s41586-023-05695-4} {\bibfield
  {journal} {\bibinfo  {journal} {Nature}\ }\textbf {\bibinfo {volume} {615}},\
  \bibinfo {pages} {231} (\bibinfo {year} {2023})}\BibitemShut {NoStop}%
\bibitem [{\citenamefont {Fläschner}\ \emph {et~al.}(2018)\citenamefont
  {Fläschner}, \citenamefont {Vogel}, \citenamefont {Tarnowski}, \citenamefont
  {Rem}, \citenamefont {Lühmann}, \citenamefont {Heyl}, \citenamefont
  {Budich}, \citenamefont {Mathey}, \citenamefont {Sengstock},\ and\
  \citenamefont {Weitenberg}}]{RN1693}%
  \BibitemOpen
  \bibfield  {author} {\bibinfo {author} {\bibfnamefont {N.}~\bibnamefont
  {Fläschner}}, \bibinfo {author} {\bibfnamefont {D.}~\bibnamefont {Vogel}},
  \bibinfo {author} {\bibfnamefont {M.}~\bibnamefont {Tarnowski}}, \bibinfo
  {author} {\bibfnamefont {B.~S.}\ \bibnamefont {Rem}}, \bibinfo {author}
  {\bibfnamefont {D.~S.}\ \bibnamefont {Lühmann}}, \bibinfo {author}
  {\bibfnamefont {M.}~\bibnamefont {Heyl}}, \bibinfo {author} {\bibfnamefont
  {J.~C.}\ \bibnamefont {Budich}}, \bibinfo {author} {\bibfnamefont
  {L.}~\bibnamefont {Mathey}}, \bibinfo {author} {\bibfnamefont
  {K.}~\bibnamefont {Sengstock}},\ and\ \bibinfo {author} {\bibfnamefont
  {C.}~\bibnamefont {Weitenberg}},\ }\bibfield  {title} {\bibinfo {title}
  {Observation of dynamical vortices after quenches in a system with
  topology},\ }\href {https://doi.org/10.1038/s41567-017-0013-8} {\bibfield
  {journal} {\bibinfo  {journal} {Nat. Phys.}\ }\textbf {\bibinfo {volume}
  {14}},\ \bibinfo {pages} {265} (\bibinfo {year} {2018})}\BibitemShut
  {NoStop}%
\bibitem [{\citenamefont {Braun}\ \emph {et~al.}(2024)\citenamefont {Braun},
  \citenamefont {Saint-Jalm}, \citenamefont {Hesse}, \citenamefont {Arceri},
  \citenamefont {Bloch},\ and\ \citenamefont {Aidelsburger}}]{RN1952}%
  \BibitemOpen
  \bibfield  {author} {\bibinfo {author} {\bibfnamefont {C.}~\bibnamefont
  {Braun}}, \bibinfo {author} {\bibfnamefont {R.}~\bibnamefont {Saint-Jalm}},
  \bibinfo {author} {\bibfnamefont {A.}~\bibnamefont {Hesse}}, \bibinfo
  {author} {\bibfnamefont {J.}~\bibnamefont {Arceri}}, \bibinfo {author}
  {\bibfnamefont {I.}~\bibnamefont {Bloch}},\ and\ \bibinfo {author}
  {\bibfnamefont {M.}~\bibnamefont {Aidelsburger}},\ }\bibfield  {title}
  {\bibinfo {title} {Real-space detection and manipulation of topological edge
  modes with ultracold atoms},\ }\href
  {https://doi.org/10.1038/s41567-024-02506-z} {\bibfield  {journal} {\bibinfo
  {journal} {Nat. Phys.}\ }\textbf {\bibinfo {volume} {20}},\ \bibinfo {pages}
  {1306} (\bibinfo {year} {2024})}\BibitemShut {NoStop}%
\bibitem [{\citenamefont {Miao}\ \emph {et~al.}(2022)\citenamefont {Miao},
  \citenamefont {Zhang}, \citenamefont {Zhao}, \citenamefont {Zhao},
  \citenamefont {Wang},\ and\ \citenamefont {Hu}}]{RN1953}%
  \BibitemOpen
  \bibfield  {author} {\bibinfo {author} {\bibfnamefont {S.}~\bibnamefont
  {Miao}}, \bibinfo {author} {\bibfnamefont {Z.}~\bibnamefont {Zhang}},
  \bibinfo {author} {\bibfnamefont {Y.}~\bibnamefont {Zhao}}, \bibinfo {author}
  {\bibfnamefont {Z.}~\bibnamefont {Zhao}}, \bibinfo {author} {\bibfnamefont
  {H.}~\bibnamefont {Wang}},\ and\ \bibinfo {author} {\bibfnamefont
  {J.}~\bibnamefont {Hu}},\ }\bibfield  {title} {\bibinfo {title} {Bosonic
  fractional quantum hall conductance in shaken honeycomb optical lattices
  without flat bands},\ }\href {https://doi.org/10.1103/PhysRevB.106.054310}
  {\bibfield  {journal} {\bibinfo  {journal} {Phys. Rev. B}\ }\textbf {\bibinfo
  {volume} {106}},\ \bibinfo {pages} {054310} (\bibinfo {year}
  {2022})}\BibitemShut {NoStop}%
\bibitem [{\citenamefont {Essam}\ and\ \citenamefont {Fisher}(1970)}]{RN2076}%
  \BibitemOpen
  \bibfield  {author} {\bibinfo {author} {\bibfnamefont {J.~W.}\ \bibnamefont
  {Essam}}\ and\ \bibinfo {author} {\bibfnamefont {M.~E.}\ \bibnamefont
  {Fisher}},\ }\bibfield  {title} {\bibinfo {title} {Some basic definitions in
  graph theory},\ }\href {https://doi.org/10.1103/RevModPhys.42.272} {\bibfield
   {journal} {\bibinfo  {journal} {Rev. Mod. Phys.}\ }\textbf {\bibinfo
  {volume} {42}},\ \bibinfo {pages} {272} (\bibinfo {year} {1970})}\BibitemShut
  {NoStop}%
\bibitem [{\citenamefont {Bradlyn}\ \emph {et~al.}(2017)\citenamefont
  {Bradlyn}, \citenamefont {Elcoro}, \citenamefont {Cano}, \citenamefont
  {Vergniory}, \citenamefont {Wang}, \citenamefont {Felser}, \citenamefont
  {Aroyo},\ and\ \citenamefont {Bernevig}}]{RN2075}%
  \BibitemOpen
  \bibfield  {author} {\bibinfo {author} {\bibfnamefont {B.}~\bibnamefont
  {Bradlyn}}, \bibinfo {author} {\bibfnamefont {L.}~\bibnamefont {Elcoro}},
  \bibinfo {author} {\bibfnamefont {J.}~\bibnamefont {Cano}}, \bibinfo {author}
  {\bibfnamefont {M.~G.}\ \bibnamefont {Vergniory}}, \bibinfo {author}
  {\bibfnamefont {Z.}~\bibnamefont {Wang}}, \bibinfo {author} {\bibfnamefont
  {C.}~\bibnamefont {Felser}}, \bibinfo {author} {\bibfnamefont {M.~I.}\
  \bibnamefont {Aroyo}},\ and\ \bibinfo {author} {\bibfnamefont {B.~A.}\
  \bibnamefont {Bernevig}},\ }\bibfield  {title} {\bibinfo {title} {Topological
  quantum chemistry},\ }\href {https://doi.org/10.1038/nature23268} {\bibfield
  {journal} {\bibinfo  {journal} {Nature}\ }\textbf {\bibinfo {volume} {547}},\
  \bibinfo {pages} {298} (\bibinfo {year} {2017})}\BibitemShut {NoStop}%
\bibitem [{\citenamefont {Alicea}\ \emph {et~al.}(2005)\citenamefont {Alicea},
  \citenamefont {Motrunich}, \citenamefont {Hermele},\ and\ \citenamefont
  {Fisher}}]{RN1997}%
  \BibitemOpen
  \bibfield  {author} {\bibinfo {author} {\bibfnamefont {J.}~\bibnamefont
  {Alicea}}, \bibinfo {author} {\bibfnamefont {O.~I.}\ \bibnamefont
  {Motrunich}}, \bibinfo {author} {\bibfnamefont {M.}~\bibnamefont {Hermele}},\
  and\ \bibinfo {author} {\bibfnamefont {M.~P.~A.}\ \bibnamefont {Fisher}},\
  }\bibfield  {title} {\bibinfo {title} {Criticality in quantum triangular
  antiferromagnets via fermionized vortices},\ }\href
  {https://doi.org/10.1103/PhysRevB.72.064407} {\bibfield  {journal} {\bibinfo
  {journal} {Phys. Rev. B}\ }\textbf {\bibinfo {volume} {72}},\ \bibinfo
  {pages} {064407} (\bibinfo {year} {2005})}\BibitemShut {NoStop}%
\bibitem [{\citenamefont {Ori}\ \emph {et~al.}(2011)\citenamefont {Ori},
  \citenamefont {Cataldo},\ and\ \citenamefont {Putz}}]{RN1996}%
  \BibitemOpen
  \bibfield  {author} {\bibinfo {author} {\bibfnamefont {O.}~\bibnamefont
  {Ori}}, \bibinfo {author} {\bibfnamefont {F.}~\bibnamefont {Cataldo}},\ and\
  \bibinfo {author} {\bibfnamefont {M.~V.}\ \bibnamefont {Putz}},\ }\bibfield
  {title} {\bibinfo {title} {Topological anisotropy of stone-wales waves in
  graphenic fragments},\ }\href {https://doi.org/10.3390/ijms12117934}
  {\bibfield  {journal} {\bibinfo  {journal} {Int. J. Mol. Sci.}\ }\textbf
  {\bibinfo {volume} {12}},\ \bibinfo {pages} {7934} (\bibinfo {year}
  {2011})}\BibitemShut {NoStop}%
\bibitem [{\citenamefont {Hecht}(2001)}]{Optics}%
  \BibitemOpen
  \bibfield  {author} {\bibinfo {author} {\bibfnamefont {E.}~\bibnamefont
  {Hecht}},\ }\href@noop {} {\emph {\bibinfo {title} {Optics}}}\ (\bibinfo
  {publisher} {Pearson Education Limited},\ \bibinfo {year} {2001})\BibitemShut
  {NoStop}%
\bibitem [{\citenamefont {Born}\ and\ \citenamefont
  {Wolf}(1999)}]{PrinciplesOfOptics}%
  \BibitemOpen
  \bibfield  {author} {\bibinfo {author} {\bibfnamefont {M.}~\bibnamefont
  {Born}}\ and\ \bibinfo {author} {\bibfnamefont {E.}~\bibnamefont {Wolf}},\
  }\href@noop {} {\emph {\bibinfo {title} {Principles of Optics}}}\ (\bibinfo
  {publisher} {Cambridge University Press},\ \bibinfo {year}
  {1999})\BibitemShut {NoStop}%
\bibitem [{\citenamefont {Steck}(2007)}]{RN542}%
  \BibitemOpen
  \bibfield  {author} {\bibinfo {author} {\bibfnamefont {D.}~\bibnamefont
  {Steck}},\ }\href@noop {} {\emph {\bibinfo {title} {Quantum and Atom
  Optics}}}\ (\bibinfo {year} {2007})\BibitemShut {NoStop}%
\bibitem [{\citenamefont {Gould}\ \emph {et~al.}(1986)\citenamefont {Gould},
  \citenamefont {Ruff},\ and\ \citenamefont {Pritchard}}]{RN730}%
  \BibitemOpen
  \bibfield  {author} {\bibinfo {author} {\bibfnamefont {P.~L.}\ \bibnamefont
  {Gould}}, \bibinfo {author} {\bibfnamefont {G.~A.}\ \bibnamefont {Ruff}},\
  and\ \bibinfo {author} {\bibfnamefont {D.~E.}\ \bibnamefont {Pritchard}},\
  }\bibfield  {title} {\bibinfo {title} {Diffraction of atoms by light: The
  near-resonant {Kapitza-Dirac} effect},\ }\href
  {https://doi.org/10.1103/PhysRevLett.56.827} {\bibfield  {journal} {\bibinfo
  {journal} {Phys. Rev. Lett.}\ }\textbf {\bibinfo {volume} {56}},\ \bibinfo
  {pages} {827} (\bibinfo {year} {1986})}\BibitemShut {NoStop}%
\bibitem [{\citenamefont {Ovchinnikov}\ \emph {et~al.}(1999)\citenamefont
  {Ovchinnikov}, \citenamefont {Müller}, \citenamefont {Doery}, \citenamefont
  {Vredenbregt}, \citenamefont {Helmerson}, \citenamefont {Rolston},\ and\
  \citenamefont {Phillips}}]{RN1926}%
  \BibitemOpen
  \bibfield  {author} {\bibinfo {author} {\bibfnamefont {Y.~B.}\ \bibnamefont
  {Ovchinnikov}}, \bibinfo {author} {\bibfnamefont {J.~H.}\ \bibnamefont
  {Müller}}, \bibinfo {author} {\bibfnamefont {M.~R.}\ \bibnamefont {Doery}},
  \bibinfo {author} {\bibfnamefont {E.~J.~D.}\ \bibnamefont {Vredenbregt}},
  \bibinfo {author} {\bibfnamefont {K.}~\bibnamefont {Helmerson}}, \bibinfo
  {author} {\bibfnamefont {S.~L.}\ \bibnamefont {Rolston}},\ and\ \bibinfo
  {author} {\bibfnamefont {W.~D.}\ \bibnamefont {Phillips}},\ }\bibfield
  {title} {\bibinfo {title} {Diffraction of a released {Bose-Einstein}
  condensate by a pulsed standing light wave},\ }\href
  {https://doi.org/10.1103/PhysRevLett.83.284} {\bibfield  {journal} {\bibinfo
  {journal} {Phys. Rev. Lett.}\ }\textbf {\bibinfo {volume} {83}},\ \bibinfo
  {pages} {284} (\bibinfo {year} {1999})}\BibitemShut {NoStop}%
\bibitem [{\citenamefont {Wen}\ \emph {et~al.}(2020)\citenamefont {Wen},
  \citenamefont {Meng}, \citenamefont {Wang}, \citenamefont {Wang},
  \citenamefont {Chen}, \citenamefont {Huang}, \citenamefont {Zhou},
  \citenamefont {Cui},\ and\ \citenamefont {Zhang}}]{RN905}%
  \BibitemOpen
  \bibfield  {author} {\bibinfo {author} {\bibfnamefont {K.}~\bibnamefont
  {Wen}}, \bibinfo {author} {\bibfnamefont {Z.}~\bibnamefont {Meng}}, \bibinfo
  {author} {\bibfnamefont {P.}~\bibnamefont {Wang}}, \bibinfo {author}
  {\bibfnamefont {L.}~\bibnamefont {Wang}}, \bibinfo {author} {\bibfnamefont
  {L.}~\bibnamefont {Chen}}, \bibinfo {author} {\bibfnamefont {L.}~\bibnamefont
  {Huang}}, \bibinfo {author} {\bibfnamefont {L.}~\bibnamefont {Zhou}},
  \bibinfo {author} {\bibfnamefont {X.}~\bibnamefont {Cui}},\ and\ \bibinfo
  {author} {\bibfnamefont {J.}~\bibnamefont {Zhang}},\ }\bibfield  {title}
  {\bibinfo {title} {Observation of sub-wavelength phase structure of matter
  wave with two-dimensional optical lattice by {Kapitza-Dirac} diffraction},\
  }\href {https://doi.org/10.1038/s41598-020-62551-5} {\bibfield  {journal}
  {\bibinfo  {journal} {Sci. Rep.}\ }\textbf {\bibinfo {volume} {10}},\
  \bibinfo {pages} {5870} (\bibinfo {year} {2020})}\BibitemShut {NoStop}%
\bibitem [{\citenamefont {Ghiglia}\ and\ \citenamefont {Pritt}(1998)}]{RN2080}%
  \BibitemOpen
  \bibfield  {author} {\bibinfo {author} {\bibfnamefont {D.~C.}\ \bibnamefont
  {Ghiglia}}\ and\ \bibinfo {author} {\bibfnamefont {M.~D.}\ \bibnamefont
  {Pritt}},\ }\href@noop {} {\emph {\bibinfo {title} {Two-Dimensional Phase
  Unwrapping: Theory, Algorithms, and Software}}}\ (\bibinfo  {publisher}
  {Wiley-Interscience},\ \bibinfo {year} {1998})\BibitemShut {NoStop}%
\bibitem [{\citenamefont {Huang}\ \emph {et~al.}(2011)\citenamefont {Huang},
  \citenamefont {Harder}, \citenamefont {Xiong}, \citenamefont {Shi},\ and\
  \citenamefont {Robinson}}]{RN1972}%
  \BibitemOpen
  \bibfield  {author} {\bibinfo {author} {\bibfnamefont {X.}~\bibnamefont
  {Huang}}, \bibinfo {author} {\bibfnamefont {R.}~\bibnamefont {Harder}},
  \bibinfo {author} {\bibfnamefont {G.}~\bibnamefont {Xiong}}, \bibinfo
  {author} {\bibfnamefont {X.}~\bibnamefont {Shi}},\ and\ \bibinfo {author}
  {\bibfnamefont {I.}~\bibnamefont {Robinson}},\ }\bibfield  {title} {\bibinfo
  {title} {Propagation uniqueness in three-dimensional coherent diffractive
  imaging},\ }\href {https://doi.org/10.1103/PhysRevB.83.224109} {\bibfield
  {journal} {\bibinfo  {journal} {Phys. Rev. B}\ }\textbf {\bibinfo {volume}
  {83}},\ \bibinfo {pages} {224109} (\bibinfo {year} {2011})}\BibitemShut
  {NoStop}%
\bibitem [{\citenamefont {Newton}\ \emph {et~al.}(2010)\citenamefont {Newton},
  \citenamefont {Harder}, \citenamefont {Huang}, \citenamefont {Xiong},\ and\
  \citenamefont {Robinson}}]{RN1973}%
  \BibitemOpen
  \bibfield  {author} {\bibinfo {author} {\bibfnamefont {M.~C.}\ \bibnamefont
  {Newton}}, \bibinfo {author} {\bibfnamefont {R.}~\bibnamefont {Harder}},
  \bibinfo {author} {\bibfnamefont {X.}~\bibnamefont {Huang}}, \bibinfo
  {author} {\bibfnamefont {G.}~\bibnamefont {Xiong}},\ and\ \bibinfo {author}
  {\bibfnamefont {I.~K.}\ \bibnamefont {Robinson}},\ }\bibfield  {title}
  {\bibinfo {title} {Phase retrieval of diffraction from highly strained
  crystals},\ }\href {https://doi.org/10.1103/PhysRevB.82.165436} {\bibfield
  {journal} {\bibinfo  {journal} {Phys. Rev. B}\ }\textbf {\bibinfo {volume}
  {82}},\ \bibinfo {pages} {165436} (\bibinfo {year} {2010})}\BibitemShut
  {NoStop}%
\bibitem [{\citenamefont {Ramos}\ \emph {et~al.}(2019)\citenamefont {Ramos},
  \citenamefont {Grønager}, \citenamefont {Andersen},\ and\ \citenamefont
  {Andreasen}}]{RN1974}%
  \BibitemOpen
  \bibfield  {author} {\bibinfo {author} {\bibfnamefont {T.}~\bibnamefont
  {Ramos}}, \bibinfo {author} {\bibfnamefont {B.~E.}\ \bibnamefont
  {Grønager}}, \bibinfo {author} {\bibfnamefont {M.~S.}\ \bibnamefont
  {Andersen}},\ and\ \bibinfo {author} {\bibfnamefont {J.~W.}\ \bibnamefont
  {Andreasen}},\ }\bibfield  {title} {\bibinfo {title} {Direct
  three-dimensional tomographic reconstruction and phase retrieval of far-field
  coherent diffraction patterns},\ }\href
  {https://doi.org/10.1103/PhysRevA.99.023801} {\bibfield  {journal} {\bibinfo
  {journal} {Phys. Rev. A}\ }\textbf {\bibinfo {volume} {99}},\ \bibinfo
  {pages} {023801} (\bibinfo {year} {2019})}\BibitemShut {NoStop}%
\bibitem [{\citenamefont {Thomas}\ \emph {et~al.}(2016)\citenamefont {Thomas},
  \citenamefont {Barter}, \citenamefont {Leung}, \citenamefont {Daiss},\ and\
  \citenamefont {Stamper-Kurn}}]{RN1907}%
  \BibitemOpen
  \bibfield  {author} {\bibinfo {author} {\bibfnamefont {C.~K.}\ \bibnamefont
  {Thomas}}, \bibinfo {author} {\bibfnamefont {T.~H.}\ \bibnamefont {Barter}},
  \bibinfo {author} {\bibfnamefont {T.~H.}\ \bibnamefont {Leung}}, \bibinfo
  {author} {\bibfnamefont {S.}~\bibnamefont {Daiss}},\ and\ \bibinfo {author}
  {\bibfnamefont {D.~M.}\ \bibnamefont {Stamper-Kurn}},\ }\bibfield  {title}
  {\bibinfo {title} {Signatures of spatial inversion asymmetry of an optical
  lattice observed in matter-wave diffraction},\ }\href
  {https://doi.org/10.1103/PhysRevA.93.063613} {\bibfield  {journal} {\bibinfo
  {journal} {Phys. Rev. A}\ }\textbf {\bibinfo {volume} {93}},\ \bibinfo
  {pages} {063613} (\bibinfo {year} {2016})}\BibitemShut {NoStop}%
\bibitem [{\citenamefont {Wen}\ \emph {et~al.}(2021)\citenamefont {Wen},
  \citenamefont {Meng}, \citenamefont {Wang}, \citenamefont {Chen},
  \citenamefont {Huang}, \citenamefont {Wang},\ and\ \citenamefont
  {Zhang}}]{RN1308}%
  \BibitemOpen
  \bibfield  {author} {\bibinfo {author} {\bibfnamefont {K.}~\bibnamefont
  {Wen}}, \bibinfo {author} {\bibfnamefont {Z.}~\bibnamefont {Meng}}, \bibinfo
  {author} {\bibfnamefont {L.}~\bibnamefont {Wang}}, \bibinfo {author}
  {\bibfnamefont {L.}~\bibnamefont {Chen}}, \bibinfo {author} {\bibfnamefont
  {L.}~\bibnamefont {Huang}}, \bibinfo {author} {\bibfnamefont
  {P.}~\bibnamefont {Wang}},\ and\ \bibinfo {author} {\bibfnamefont
  {J.}~\bibnamefont {Zhang}},\ }\bibfield  {title} {\bibinfo {title}
  {Experimental study of tune-out wavelengths for spin-dependent optical
  lattice in {$^{87}$Rb} {Bose–Einstein} condensation},\ }\href
  {https://doi.org/10.1364/JOSAB.432448} {\bibfield  {journal} {\bibinfo
  {journal} {J. Opt. Soc. Am. B}\ }\textbf {\bibinfo {volume} {38}},\ \bibinfo
  {pages} {3269} (\bibinfo {year} {2021})}\BibitemShut {NoStop}%
\end{thebibliography}%



\begin{widetext}
\setcounter{equation}{0}
\renewcommand{\theequation}{S\arabic{equation}}

\section*{Supplementary Material}

\subsection*{I. Theoretical description of diffraction dynamics}

The Hamiltonian of the system is given by
\begin{equation}
\hat{H} = \frac{\hat{\mathbf{p}}^2}{2M} + V(\mathbf{r}),
\end{equation}
where \( M \) is the atomic mass and \( V(\mathbf{r}) \) denotes the lattice potential formed by the laser beams, as described in Eq.~(1) of the main text. The eigenstates of this Hamiltonian are Bloch states \( \left\vert n,\mathbf{q} \right\rangle \), labeled by the quasimomentum \( \mathbf{q} \) and the band index \( n \), and satisfy the eigenvalue equation
\begin{equation}
\hat{H} \left\vert n,\mathbf{q} \right\rangle = E_{n,\mathbf{q}} \left\vert n,\mathbf{q} \right\rangle,
\end{equation}
with eigenenergies \( E_{n,\mathbf{q}} \). Due to the spatial periodicity, the momentum distribution of the Bloch states consists of discrete components separated by reciprocal lattice vectors $\mathbf{b}_p = l \mathbf{b}_1 + m \mathbf{b}_2$ with $p = (l, m) \in \mathbb{Z}^2$. The basis vectors are defined as $\mathbf{b}_1 = \mathbf{k}_2 - \mathbf{k}_3$ and $\mathbf{b}_2 = \mathbf{k}_3 - \mathbf{k}_1$ with \( \mathbf{k}_j \) being the wave vector of the \textit{j}th laser beam. As a result, the Bloch states can be expanded in the discrete plane-wave basis as
\begin{equation}
\left\vert n,\mathbf{q} \right\rangle = \sum_{p} a_{n,\mathbf{q}}(p)\, e^{i(\mathbf{q} + \mathbf{b}_p) \cdot \mathbf{r}},
\label{Schr1}
\end{equation}
where \( a_{n,\mathbf{q}}(p) \) are the expansion coefficients.

To model diffraction dynamics, consider a Bose–Einstein condensate (BEC) initially prepared in a plane-wave state \( \left\vert \Psi(t = 0) \right\rangle = \left\vert \phi_{\mathbf{q}} \right\rangle=e^{i\mathbf{q}\cdot \mathbf{r}}\). Upon sudden loading into the lattice, this state can be expressed as a coherent superposition of Bloch states at fixed quasimomentum \( \mathbf{q} \):
\begin{equation}
\left\vert \Psi(0) \right\rangle = \sum_{n} \left\vert n, \mathbf{q} \right\rangle \left\langle n, \mathbf{q} \middle| \phi_{\mathbf{q}} \right\rangle=\sum_{n}a_{n,\mathbf{q}}^{*}(p_0)\left\vert n, \mathbf{q} \right\rangle,
\label{Schr2}
\end{equation}
where the index $p_0 = (0, 0)$. During the subsequent evolution in the lattice for a hold time \( \tau \), the system undergoes full quantum evolution, yielding the time-dependent state
\begin{equation}
\left\vert \Psi(\tau) \right\rangle = \sum_{n} a_{n,\mathbf{q}}^{*}(p_0)e^{-i E_{n,\mathbf{q}} \tau / \hbar} \left\vert n, \mathbf{q} \right\rangle.
\label{Schr3}
\end{equation}
After the evolution, the lattice is abruptly switched off at time \( \tau \), and the wavefunction is projected onto the plane-wave measurement basis. The amplitude \( c_{\mathbf{q}}(p) \) of each momentum component is then given by
\begin{equation}
c_{\mathbf{q}}(p) = \sum_{n} a_{n,\mathbf{q}}^{*}(p_0)a_{n, \mathbf{q}}(p)\, e^{-i E_{n}(\mathbf{q}) \tau / \hbar}.
\label{Schr4}
\end{equation}
The interference between the phase factors \( e^{-i E_{n} \tau / \hbar} \) leads to diffraction patterns that evolve with the hold time \( \tau \).

\subsection*{II. Experimental scheme}

An ultracold Bose–Einstein condensate (BEC) of up to \( 5 \times 10^5 \) Rubidium-87 (\(^{87}\mathrm{Rb}\)) atoms is first prepared in the spin state \( \left\vert F = 2, m_F = 2 \right\rangle \) in a crossed optical dipole trap. The atoms are then transferred to the spin state \( \left\vert F = 1, m_F = 1 \right\rangle \) via a rapid adiabatic passage induced by a microwave (MW) transition. To continuously tune the lattice structure between dual honeycomb and hexagonal lattices, we employ three laser beams derived from a Ti:sapphire laser operating at wavelength \( \lambda \), providing a continuously tunable wavelength range of \( \Delta\lambda = 46\,\mathrm{nm} \) (from \( 774\,\mathrm{nm} \) to \( 820\,\mathrm{nm} \)). The three beams interfere at an angle of \( 120^\circ \) in the \( xy \)-plane. Their polarizations are linear in the plane of the intersecting area and perpendicular to a homogeneous magnetic guiding field of $B_0=2.7\,$G along the $z$ direction, which defines the quantization axis.

When placed in the optical lattice, the atoms experience the ac Stark effect, which manifests as a shift in their internal energy levels due to interaction with the optical field. This effect is primarily influenced by the scalar, vector, and tensor polarizabilities of the atoms. The scalar polarizability contributes uniformly to the energy shift, regardless of the internal state of the atom, while the vector polarizability is linearly dependent on the magnetic quantum number $m_F$, leading to state-dependent shifts \cite{RN1308}. The tensor polarizability vanished for our experiment owing to the large detuning of lattice beam compared to the $^{87}$Rb relevant atomic transitions. The complete lattice potential \( V(\mathbf{r}) \) in this experiment can be expressed as the sum of the scalar Stark shift potential \( V_{\mathrm{s}}(\mathbf{r}) \) and the vector Stark shift potential \( V_{\mathrm{v}}(\mathbf{r}) \), as given by Eq.~(1) of the main text. The scalar polarizability is defined as $\alpha^{(0)} = \frac{\pi \epsilon_0 c^{3} \Gamma_{D_2}}{\omega_0^3} \left( \frac{1}{\omega_{D_1} - \omega} + \frac{2}{\omega_{D_2} - \omega} \right)$, and the vector polarizability is given by
$\alpha^{(1)} = \frac{\pi \epsilon_0 c^{3} \Gamma_{D_2}}{\omega_0^3} \left( \frac{1}{\omega_{D_1} - \omega} - \frac{1}{\omega_{D_2} - \omega} \right) F g_F$, where \( F \) is the total atomic angular momentum of the hyperfine ground state. The frequencies \(\omega_{D_1}\) and \(\omega_{D_2}\) correspond to the resonance transitions at wavelengths 795 nm and 780 nm, respectively. The effective average transition frequency is given by \(\omega_0 = \frac{1}{3} \omega_{D_1} + \frac{2}{3} \omega_{D_2}\). \(\Gamma_{D_2}\) is the natural linewidth of the \(D_2\) transition, and \(g_F\) is the hyperfine Land\'{e} \(g\)-factor.

\setcounter{figure}{0}
\renewcommand{\thefigure}{S\arabic{figure}}
\renewcommand{\figurename}{FIG}
\begin{figure}[tb]
\includegraphics[width=4in]{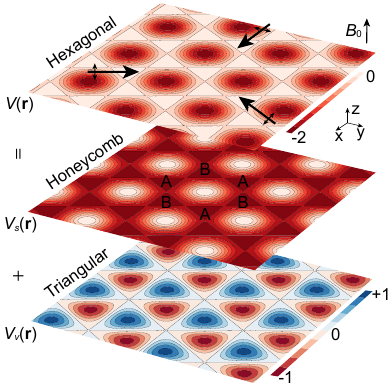}
\caption{Laser geometry and lattice configuration with $\alpha^{(0)}=\alpha^{(1)}>0$ for atoms in the spin state $|F=1,m_F=1\rangle$. The scalar light shift with \( \alpha^{(0)} > 0 \) generates a honeycomb lattice potential \( V_\mathrm{s}(\mathbf{r}) \), while the vector light shift gives rise to a triangular lattice potential \( V_\mathrm{v}(\mathbf{r}) \). An equal-weight superposition of the scalar and vector potentials with $\alpha^{(0)} = \alpha^{(1)} > 0$ results in a hexagonal lattice potential \( V(\mathbf{r}) \).}\label{SFig1}
\end{figure}

The lattice geometry is strongly influenced by the tunable polarizability ratio \( \alpha = \alpha^{(0)} / \alpha^{(1)} \), which is controlled via the laser wavelength. Specifically, when \( \alpha = 1 \), the lattice potential \( V(\mathbf{r}) \) forms a hexagonal lattice, as shown in Fig.~\ref{SFig1}, whereas for \( \alpha = -1 \), it forms a honeycomb lattice. The resulting lattice geometries and their associated symmetries are also reflected in the topological band properties, as illustrated in Fig.~\ref{SFig2}(a). The honeycomb lattice, possessing \( C_6 \) rotational symmetry, hosts a Dirac point at the \(K\) point between the first and second energy bands. When \( \alpha \) deviates from \(-1\), the \( C_6 \) symmetry is reduced to \( C_3 \), resulting in the opening of a bandgap at the Dirac point. Similarly, the hexagonal lattice also features a Dirac-like degeneracy at the \(K\) point between the second and third bands, which is protected by the same \( C_6 \) rotational symmetry. When \( \alpha \) deviates from \(+1\), the reduction of symmetry from \( C_6 \) to \( C_3 \) likewise leads to a bandgap opening. Interestingly, the triangular lattice formed by the pure vector potential hosts a threefold degeneracy point at the \(\Gamma\) point, as shown in the third panel of Fig.~\ref{SFig2}(a). In addition, the reduced \( C_3 \) rotational symmetry is clearly manifested in the triangular diffraction patterns, induced by phase wrapping in the strong-pulse Raman--Nath regime, as shown in Figs.~\ref{SFig2}(b)-\ref{SFig2}(d).

\begin{figure}[tb]
\includegraphics[width=4in]{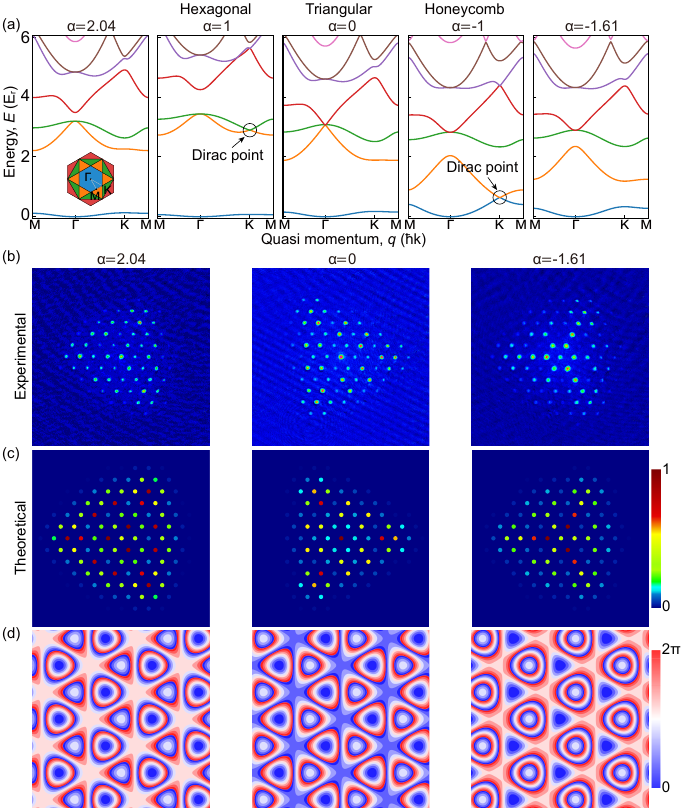}
\caption{(a) Energy bands for different lattice structures with the lattice depth $V_0=8 \, E_r$. (b), (c) Kapitza--Dirac diffraction of a BEC in lattices with \( C_3 \) rotational symmetry under the strong-pulse Raman--Nath regime. Panel (b) shows the experimental result, while (c) presents the corresponding theoretical simulation. The pulse strength and duration are \( V_0 = 95\,E_r \) and \( \tau = 7~\mu\text{s} \), corresponding to phase wrapping with \( V_0 \tau / \hbar = 2.47 \times 2\pi \). (d) Schematic of phase wrapping with \( V_0 \tau / \hbar = 2.47 \times 2\pi \).}\label{SFig2}
\end{figure}

By tuning the wavelength of the lattice beams to adjust the polarizability ratio $\alpha$, we can also achieve continuous control over the lattice structure for atoms in the spin state $\left\vert F=2, m_F=2\right\rangle$ , as shown in Fig.~\ref{SFig3}. A honeycomb lattice is realized at $\lambda_1=785.09$nm with $\alpha=-1$, whereas a hexagonal lattice forms at $\lambda_1=794.97$nm with $\alpha=1$. In the latter case, however, measurement is limited by heating effects arising from near-resonant light-atom interactions.

\begin{figure}[tb]
\includegraphics[width=3.6in]{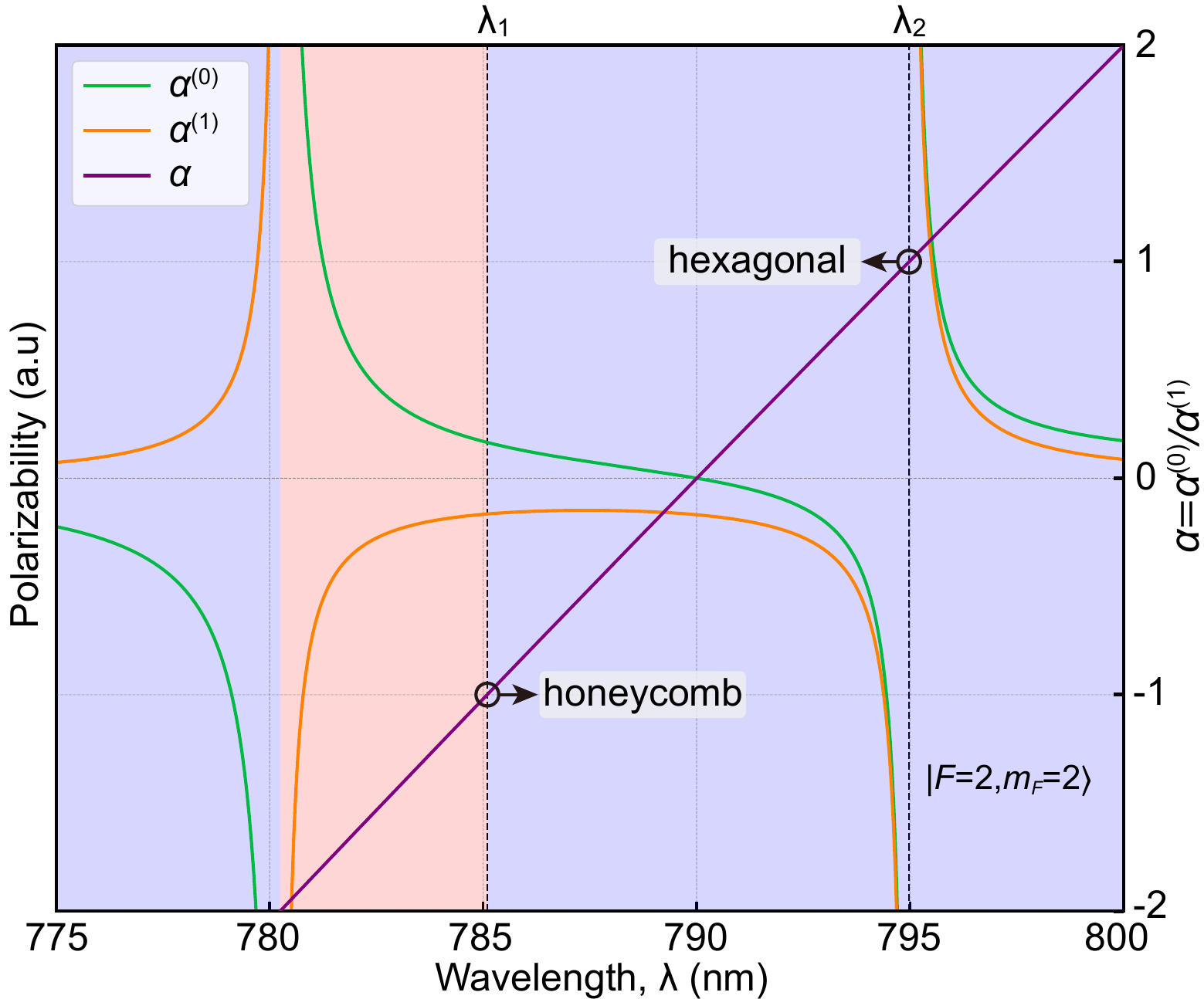}
\caption{The polarizability as a function of wavelength for $^{87}$Rb atoms in the hyperfine ground state $\left\vert F = 2, m_F = 2 \right\rangle$. The ratio of scalar to vector polarizability, with $\alpha = -1$ at $\lambda_1 = 785.09\,\text{nm}$ and $\alpha = 1$ at $\lambda_2=794.97\,\text{nm}$, corresponds to the honeycomb and hexagonal lattices, respectively.}
\label{SFig3}
\end{figure}

\subsection*{III. Comparison between Bragg and Kapitza--Dirac Diffraction}

Bragg diffraction occurs under energy–momentum resonance conditions, enabling atoms to be selectively transferred into a single momentum state. In contrast, Kapitza–Dirac (KD) diffraction occurs in the far-off-resonant regime, generating multiple symmetric momentum components whose populations follow Bessel-function distributions. Table~\ref{Table:Bragg_KD_diffraction} summarizes the key differences between Bragg and KD diffraction, along with the characteristic features of various KD regimes in the context of atom–light interactions within standing-wave optical lattices.

\setcounter{figure}{0}
\renewcommand{\thefigure}{\arabic{figure}}
\renewcommand{\figurename}{Table}
\begin{figure}[hb]
\centering
\includegraphics[width=6.5in]{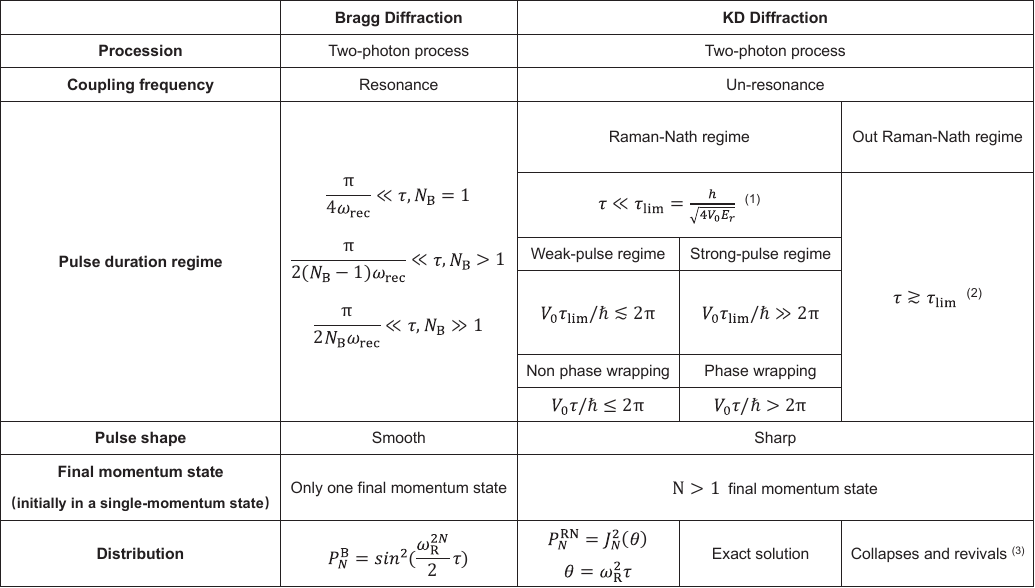}
\caption{
Key differences between Bragg and Kapitza--Dirac diffraction, along with the characteristic features of various KD regimes. $\omega_{\mathrm{rec}}$ represents the frequency corresponding to the recoil energy, $N_B$ indicates the Nth-order Bragg diffraction, $\tau$ is the interaction time, $\Omega_0$ is the Rabi frequency of the single-photon process, $P_N^B$ refers to the population probability at the Nth-order Bragg diffraction point, $P_N^{RN}$ denotes the population probability at the Nth-order KD diffraction point, $J_N(\theta)$ represents the Nth-order Bessel function, $\omega_R^{2N}$ corresponds to the Rabi frequency of the 2N-photon process. (1) The atomic kinetic energy is neglected, and the standing wave acts as a thin phase grating. (2) Non-negligible atomic kinetic energy. (3) Solving the Schrödinger equation exactly.}\label{Table:Bragg_KD_diffraction}
\end{figure}

\end{widetext}

\end{document}